\documentclass[ twocolumn]{aastex62}

 \usepackage{amsmath}
 \usepackage{amsfonts}
\usepackage{amssymb}
 \usepackage{xcolor}
 \usepackage{soul}
 \usepackage{booktabs}
 \sethlcolor{green}

\hypersetup{linkcolor=red,citecolor=blue,filecolor=cyan,urlcolor=magenta}

\shorttitle{Bursty Star Formation}
\shortauthors{Emami et al.}
\begin{document}

%% abreviations:
\newcommand{\huv}{$L_{H\alpha}/L_{UV}$}
\newcommand{\del}{$\Delta \text{log}(L_{H\alpha})$}
\newcommand{\col}{green!65!blue} % Naj Added
\newcommand{\coll}{red!85!blue} % Removed
\newcommand{\colll}{orange!98!green} %Brian added

\title{A Closer look at Bursty Star Formation with $L_{H\alpha}$ and $L_{UV}$ Distributions}
%\title{Characterizing bursty star formation with $L_{H\alpha}/L_{UV}$ distribution; A new approach}  % textbf: bold font
\author{Najmeh Emami}
\affiliation{Department of Physics and Astronomy, University of California Riverside, Riverside, CA 92521 USA}

\author{Brian Siana}
\affiliation{Department of Physics and Astronomy, University of California Riverside, Riverside, CA 92521 USA}

\author{Daniel R. Weisz}
\affiliation{Department of Astronomy, University of California Berkeley, Berkeley, CA 94720, USA}

\author{Benjamin D. Johnson}
\affiliation{Harvard-Smithsonian Center for Astrophysics, Cambridge, MA 02138, USA}

\author{Xiangcheng Ma}
\affiliation{Department of Astronomy, University of California Berkeley, Berkeley, CA 94720, USA}

\author{Kareem El-Badry}
\affiliation{Department of Astronomy, University of California Berkeley, Berkeley, CA 94720, USA}
\begin{abstract}

We investigate the bursty star formation histories (SFHs) of dwarf galaxies using the distribution of log(\huv) of 185 local galaxies. We expand on the work of \citet{Weisz12} to consider a wider range of SFHs and stellar metallicities, and show that there are large degeneracies in a periodic, top-hat burst model. We argue that all galaxies of a given mass have similar SFHs and we can therefore include the $L_{H\alpha}$ distributions (subtracting the median trend with stellar mass, referred to as \del) in our analyses. \del\ traces the amplitude of the bursts, and log(\huv) is a function of timescale, amplitude, and shape of the bursts. We examine the 2-dimensional distribution of these two indicators to constrain the SFHs. We use exponentially rising/falling bursts to determine timescales ($e$-folding time, $\tau$). We find that galaxies below $10^{7.5}$ M$_{\odot}$ undergo large (maximum amplitudes of $\sim 100$) and rapid ($\tau < 30$ Myr) bursts, while galaxies above $10^{8.5}$ M$_{\odot}$ experience smaller (maximum amplitudes $\sim 10$), slower ($\tau \gtrsim 300$ Myr) bursts. We compare to the FIRE-2 hydrodynamical simulations and find that the burst amplitudes agree with observations, but they are too rapid in intermediate-mass galaxies ($M_* > 10^{8}$ M$_{\odot}$). Finally, we confirm that stochastic sampling of the stellar mass function can not reproduce the observed distributions unless the standard assumptions of cluster and stellar mass functions are changed. With the next generation of telescopes, measurements of $L_{UV}$ and $L_{H\alpha}$ will become available for dwarf galaxies at high-redshift, enabling similar analyses of galaxies in the early universe.% The application of this analysis on high-redshift low-mass galaxies is intriguing in the sense that it enables us to explore the burstiness at earlier times as more data of $H\alpha$ and $UV$ will be available by the launch of new generations of telescopes, JWST, GMT, WFIRST, etc. } 

\end{abstract}

\keywords{galaxies: dwarf --- galaxies: evolution --- galaxies: formation  --- galaxies: star formation}

\section{Introduction}

An active area of research in galaxy evolution is understanding ``feedback'' -- energy and/or momentum deposition into the interstellar medium -- from stars and accreting black holes. It is generally believed that star formation is suppressed in high mass galaxies by feedback from the central, supermassive black holes and in dwarf galaxies by feedback from massive stars (photoionization heating, stellar winds, radiation pressure, and supernovae) \citep{Hopkins2014,Keres_2009, Springel_2005, Governato_2010, Somerville_1999}. 

However, there are still significant uncertainties in how the various forms of feedback couple with the gas and the efficiency with which it heats or expels gas. When different sub-grid prescriptions for stellar feedback are implemented in hydrodynamical simulations, it can result in markedly different predictions of the characteristics of galaxies. One generic feature of hydrodynamical simulations of dwarf galaxies that include strong stellar feedback, is large variations in the star formation rates (SFRs), often referred to as ``bursty'' star formation. Simulations with different feedback prescriptions produce bursts of star formation with very different characteristics (e.g. amplitude and duration). Because it is in principle possible to observe large variations in SFR, observers can test these feedback prescriptions to better understand the physical mechanisms that regulate star formation in dwarf galaxies. 

The primary method by which one can measure the burstiness is to use indicators (observables) of star formation that trace different time scales. The two most common indicators are the luminosity of the (non-scattering) Hydrogen recombination lines (such as H$\alpha$ and H$\beta$), and the far-ultraviolet (far-UV) continuum (1300 \AA\ $< \lambda <$ 2000 \AA) luminosity density ($L_{UV}$). Hereafter we refer to the logarithm of the ratio of these two observables, log(\huv). $L_{H\alpha}$ is a byproduct of the ionizing radiation from short-lived O-stars. Therefore, during an episode of constant star formation, $L_{H\alpha}$ equilibrates rapidly, as the rate of O-star supernovae equals the rate of O-star formation. $L_{UV}$, on the other hand, is produced by O-stars as well as longer-lived B and A stars.  Therefore, $L_{UV}$ takes much longer to reach equilibrium after an episode of constant star formation. Using the stellar population synthesis models of \citet{Bruzual_Charlott2003} for constant star formation, we find equilibrium time scales \citep[reaching 90\% of the equilibrium value,][]{Kennicutt_2012} of 5 and 100 Myr for $L_{H\alpha}$ and $L_{UV}$, respectively. Because of this, both $L_{H\alpha}$ and $L_{UV}$ accurately trace the SFR of any galaxy whose SFR changes on time scales much larger than 100 Myr. Thus, the ratio, log(\huv), remains approximately constant. However, if the SFR changes on shorter time scales than 100 Myr, $L_{UV}$ will no longer follow the SFR and the ratio, log(\huv), will vary. Therefore, the distribution of the log(\huv) can inform whether or not the SFR of galaxies changes on time scales less than 100 Myr.

We note that these observables ($L_{H\alpha}$, $L_{UV}$) are often used to determine physical properties of galaxies like SFR and the ionizing photon production efficiency ($\xi_{ion}$) \citep{Bouwens_2015, Duncan_2015, Robertson_2013} under the assumption that star formation varies slowly with time. Therefore, understanding bursty star formation is critical for interpreting these observables \citep{Dominguez_2015}. 

There have been several analyses of the distributions of \huv\ attempting to extract information about the typical amplitudes, durations, and periods of the bursts \citep{Glazebrook1999, Iglesias_2004, lee_2011, Weisz12, Kauffmann_2014, Dominguez_2015}. Several factors other than SFH can affect the \huv, as discussed by  \citet{Iglesias_2004, Lee_2009, Meurer_2009, Boselli2009}, and \citet{2016_Guo}.
Here, we list all of these factors other than the star formation history (which is the subject of this paper) and briefly explain how they affect \huv.
\begin{itemize}
   
    \item Dust extinction: Because dust extinction is often a strong function of wavelength, and the nebular and stellar UV continuum emission can arise from stars with different spatial distributions relative to dust, the effect of dust on the observed \huv\ can be considerable \citep{Kewley_2002, Lee_2009}.
    
    \item Escape of ionizing photons: If ionizing photons are escaping from the galaxy \citep{Steidel_2001, Shapley_2006, Siana_2007},  the photoionization rate (and therefore, $L_{H\alpha}$) will be lower than expected under the assumption that all ionizing photons are absorbed in HII regions.
    
    \item Initial stellar mass function (ISMF): The \huv\ is influenced by the relative number of stars at each mass, so variations in the ISMF will affect the ratio. This can include ``effective ISMFs'', where the star-forming clouds are not sufficiently massive to fully sample the high-mass end of the ISMF \citep{Hoversten_2008, Pflamm-Altenburg2007, Pflamm-Altenburg2009}. The initial cluster mass function (ICMF) can also affect these ``effective ISMFs'' because it determines the relative number of clusters that do and do not fully sample the high mass end of the ISMF.
    
    %\item \textcolor{\col}{Initial cluster mass function (ICMF): Stars are born in star clusters and the distribution of clusters (ICMF) can also affect the relative number of massive stars to be formed, thus can change the \huv.}
    
    \item Stellar metallicity: Stars with lower metal abundances will be hotter at the same mass, resulting in a higher \huv\ \citep{Bicker_2005, Boselli2009}.
    \item Stellar models: Inclusion of binaries \citep{Eldridge_2012} and rotating stars \citep{Choi_2017} in the stellar evolution modelling will increase \huv. 
\end{itemize}
 
There are several papers discussing these effects on the \huv\ distribution, but many have concluded that the most important effects are bursty star formation and (or) variations in the IMF \citep{Meurer_2009, 2016_Guo, Mehta_2017}. The main focus of this paper is to use observable distributions to better understand bursty star formation. Specifically, we aim to: (1) more fully explore the parameter space of bursty star formation models. (2) Break degeneracies and minimize uncertainties in these models. (3) Determine typical timescales for the rise/fall of star formation as a function of galaxy stellar mass. (4) Better understand whether burstiness or IMF variations explain the observed $L_{H\alpha}/L_{UV}$ distributions.

First we describe the observational data which will be used throughout this paper in section \ref{sec:data}. In Section \ref{sec:review}, we review previous efforts to determine the burstiness parameters (amplitude, period, duration) using the \huv\ distribution. We also introduce our improved method to more completely explore the parameter space and compare with previous results. In Section \ref{sec:lha}, we propose combining the $L_{H\alpha}$ distribution with the \huv\ distribution to better constrain bursty star formation models. In Section \ref{sec:exp}, we introduce a new exponential burst model to better constrain the timescales for the rise and fall of SFRs in dwarf galaxies. In Section \ref{sec:discussion}, we compare to predictions from hydrodynamical simulations and discuss the physical implications. We also examine the effect of stochastic IMF sampling on our analysis and discuss the results in this section as well as the effects of escape fraction and dust attenuation on the observed distribution.

\section{Observational Data}
\label{sec:data}

In this work, we use the same far-ultraviolet ($FUV$) and $H\alpha$ photometry as W12 from the 11 Mpc $H\alpha$ and UV Galaxy Survey \citep[11HUGS, ][]{Kennicutt_2008, Lee_2009}. The primary sample is complete in including all nearby galaxies within 11 Mpc  and consists of spirals and irregulars that avoid the Galactic plane($|b| > 20^{\circ}$) and are brighter than $B = 15$ mag. Stellar masses were determined using optical photometry from the Sloan Digital Sky Survey and $Spitzer$ mid-IR (IRAC) photometry from the Local Volume Legacy (LVL) Survey. 
Thus the W12 sample is a subsample of 11Hugs galaxies for which their optical and IR measurements are available.
The data are corrected for both Galactic foreground dust extinction and extinction within the target galaxy. See W12 and \citet{Lee_2009} for further details and sample completeness. 
We decided to remove five galaxies with either log(\huv) $< -3.4$ or log(\huv) $> -1.8$ from the sample. Two of the outliers (UGCA281, MRK475) are Wolf-Rayet galaxies. The other three outliers have extremely high or low log(\huv) such that the stellar synthesis models are not able to reproduce them (UGCA438, KDG61, UGC7408).

\section{Review of the Methods}
\label{sec:review}

Below, we discuss the methods used in previous studies and argue the strengths and shortcomings of those methods.

\cite{Iglesias_2004} and \cite{Boselli2009} considered instantaneous bursts of SF superimposed on a baseline with a few different time intervals between bursts. \cite{Meurer_2009} used a SF model of Gaussian bursts added (or ``gasps'' subtracted) from a constant SFR, in which the FWHM of the Gaussians represents the duration, and also affects timescale for fractional changes in SFR. All three studies assumed discrete model parameters and did not fully and systematically explore the parameter space. It is noteworthy that \cite{Boselli2009} was the first to introduce different model parameters for galaxies of different stellar masses, pointing out that low mass galaxies show a larger spread in the log(\huv) distribution. They conclude that this larger spread in the log(\huv) distribution is due to sporadic bursts of SF (with periods of 10 Myr) whereas the small spread in the log(\huv) distribution of massive galaxies suggests a roughly constant SFR.

\citet{Weisz12} (hereafter W12) made a notable step forward and divided their observed sample into five stellar mass bins from $~10^{6}$ to $~10^{11} M_{\odot}$ and assumed that all galaxies of similar stellar masses have star formation histories with the same parameters and could be considered random samples in time of each SFH. They defined their SF models as top-hat, periodic bursts superimposed on a constant baseline. They then determined the log(\huv) distribution from models with different burst periods (P, the time interval between two consecutive bursts), durations (D, the time length when the star formation is in burst), and amplitudes (A, the SFR at burst relative to the baseline SFR) and used a Kolmogorov-Smirnov test to identify the parameters that best reproduce the observed log(\huv) distribution in each mass bin. The other improvement in W12 was the finer sampling of parameter space. The main conclusion of W12 is that galaxies with the lowest stellar masses have higher amplitude bursts (A $\sim 30\times$ the baseline rate), relatively long durations (D $\sim 30-40$ Myr), and long periods (P = $250$ Myr). The highest mass bins are characterized by almost constant SFRs with an occasional modest burst favoring SF models with short duration (D $\sim$ 6 Myr) and modest amplitudes (A $\sim 10$).

In the Appendix, we describe an analysis similar to that of W12, but with three significant improvements to better determine the best parameters of bursty SF. First, we use appropriate (lower) stellar metallicities for the lower mass galaxies, as W12 used solar metallicities for all galaxies.  This should change the predicted log(\huv) distributions as stars of the same mass at lower metallicity will be hotter and have a larger log(\huv) ratio. Second, we expand the parameter space, as some of the best-fit parameters in W12 were at the edge of the explored parameter space. Finally, we adopt a probabilistic approach to determine the best SF history parameters. This allows us to fully explore the parameter space, determine the relative merit (the likelihood) of each set of parameters, and search for any significant degeneracies between the parameters. 

\begin{figure*}
\begin{center} 
\includegraphics[width=1\linewidth]{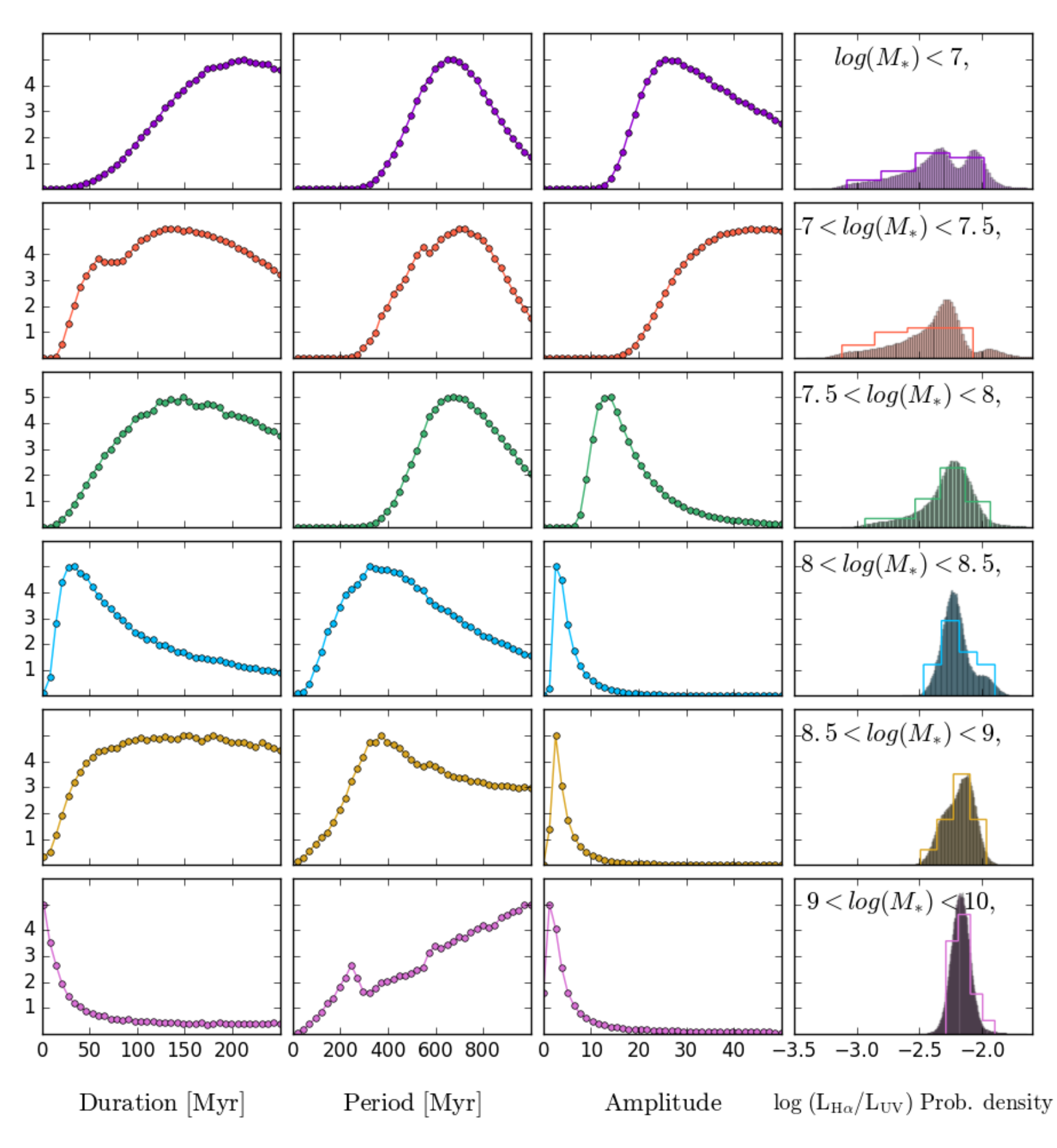} 
\caption{ Likelihood function of duration, period and amplitude (relative to the baseline), as well as the log(\huv) distribution of the best fit models to the observed log(\huv) distribution for each mass bin, assuming metallicities from \cite{Andrews_2013}. The open histograms show the observed sample and the filled histograms show the distribution of the best-fit model.} 
\label{fig:11_2}
\end{center}
\end{figure*}

The results of our method are presented in Figure \ref{fig:11_2}, which shows the marginalized likelihood of the duration, period and amplitude for the six stellar mass ranges defined in Table \ref{table-1}. On the right, the observed log(\huv) distributions for each mass bin are plotted as unfilled histograms along with the distribution of the best-fit models as filled histograms.

Galaxies with $M\geq 10^8 M_{\odot}$ have best-fit amplitudes $<3$, signifying relatively stable star formation histories. However, such low amplitudes are very sensitive to the assumed errors in the observed luminosities and dust extinction corrections. On the other hand, all of the stellar mass bins at $M < 10^8 M_{\odot}$ have best-fit amplitudes $> 15$, suggesting dramatic bursts of star formation are necessary to explain the large spread of log(\huv) seen in these galaxies.

Our new analysis demonstrates that W12 did not probe the parameter space with the highest likelihood, as the posteriors peak at period of 500-900 Myr,  Figure \ref{fig:11_2} second column (where W12 analyzed P$<250$ Myr).  This is important for the duration parameter as well, as it is highly correlated with the period (see below).  Thus, the new duration estimates are considerably larger (D$\sim40-250$ Myr) compared to D$\sim20$ Myr in W12. The amplitude estimates are uncertain, but broadly agree with the values in W12. 

\begin{figure*}
\begin{center} 
\includegraphics[width=1\linewidth]{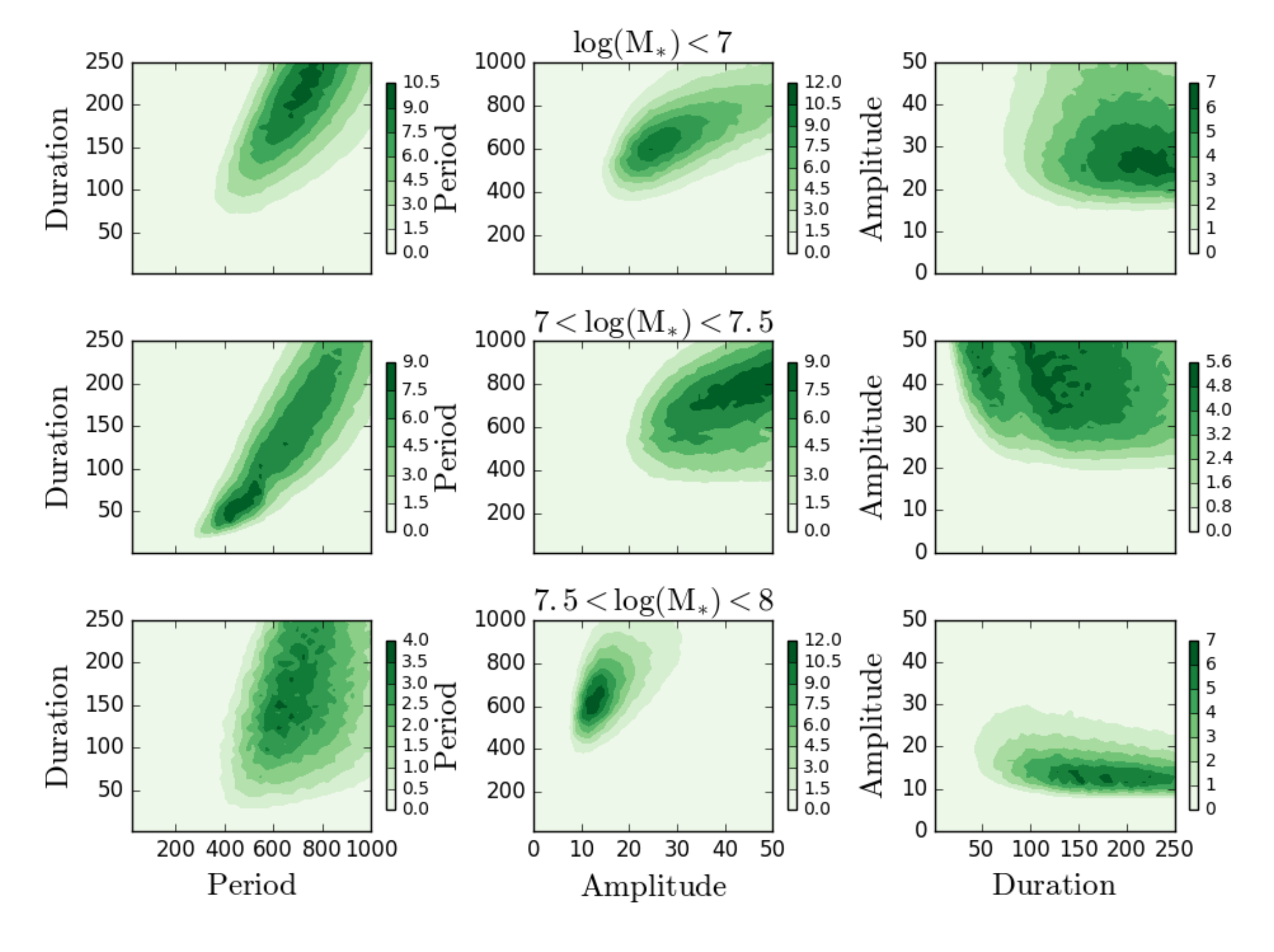}
\caption{2D  contour plots of (D,P), (P,A) and (A, D). Each row corresponds to a different mass range of $M\leq10^{7}\ M_\odot$, $10^7<M \leq 10^7.5\  M_{\odot}$ and $10^7.5<M \leq 10^8\  M_{\odot}$ from top to bottom. The probability densities are denoted as colorbars on the right of each panel. The strong degeneracy in duration and period can be seen for all mass ranges shown here.}
\label{fig:2d}
\end{center}
\end{figure*}

To better understand the degeneracies between the model parameters, we plot the 2D contours of the three lowest mass bins in Figure \ref{fig:2d}. In the Duration-Period contour plots, there is a linear degeneracy between period and duration, such that increasing the duration of the burst requires a similar increase in the period.

These degeneracies result in large uncertainties in the marginalized posteriors for both duration and period. Specifically, the burst durations of the low mass galaxies can be 50 or 250 Myr (or larger). Furthermore, we know that this periodic top-hat model is not accurate in that the star formation does not instantly change and is not truly periodic.  Therefore, it is difficult to know whether or not these best-fit parameters reflect the true values of typical duration and period in these galaxies.

\section{A New Approach to Characterizing Bursty Star Formation}

So far we reviewed the previous studies and more carefully explored the top-hat periodic SF model of W12 to determine parameters of SFHs based only on the log(\huv) distribution. 

However, two concerns ultimately arise from our new analysis.  First, our new probabilistic approach allows us to see that significant degeneracies exist between the model parameters (see Figure \ref{fig:2d}). This is perhaps not surprising, as there are three parameters fit to a single  log(\huv) distribution. Ultimately, it is necessary to use additional observables to fit complicated star formation histories. 

Second, we know that the periodic, top-hat burst model is unphysical, as it requires instantaneous changes in SFR. Indeed, the best-fit model predicts {\it other} observable distributions that do not look like those observed. For example, the model predicts a bimodal \del\ distribution at each mass, rather than the observed distribution, where galaxies are more evenly distributed in \del, suggesting that the typical galaxy spends a significant fraction of its lifetime near the average SFR, rather than in low and high states.

In this section, we seek to add additional observables to constrain the star formation histories, and we choose a somewhat simpler parametrization in order to focus on the timescales of the transitions from burst to quiescence and back. 

\subsection{The $L_{H\alpha}$ Distribution}
\label{sec:lha}

\begin{figure}
\includegraphics[width=1\linewidth]{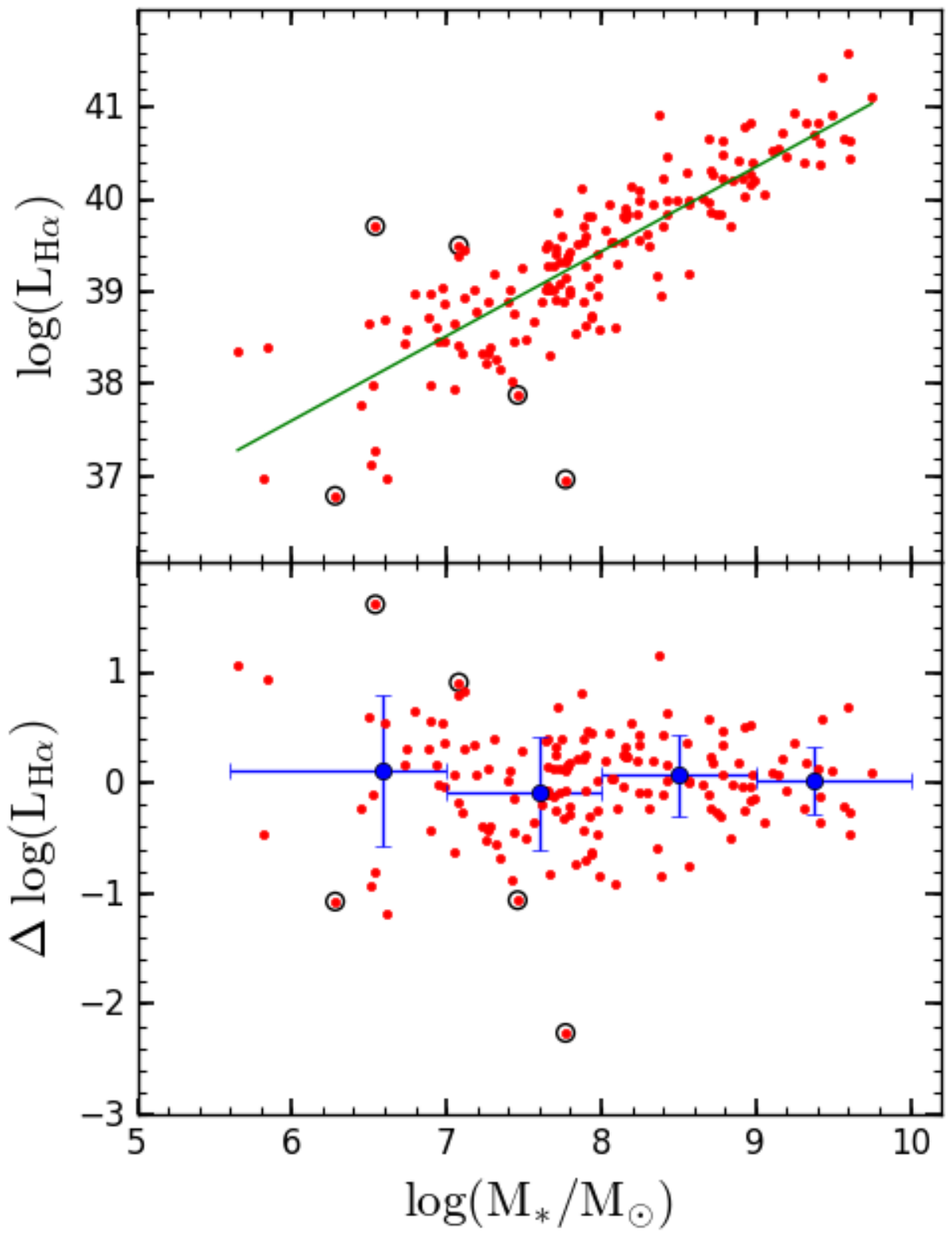}
\centering
\caption{ Top: $log(L_{H\alpha})$ vs.  $log(M_{*})$ relation for the W12 sample (red dots). The correlation between these two observables is known as the star-forming ``main sequence.'' The black circles show outliers in log(\huv) (either $>-1.8$ or $<-3.4$) that are removed from the sample for the subsequent analysis. Bottom: The deviation of the log($L_{H\alpha}$), \del, from the linear relation green line in the top panel. The blue circles are the median in each stellar mass bin. The horizontal error bars denote the stellar mass range of each bin and the vertical error bars are the standard deviation in each bin. The scatter around the mean trend (green line) increases toward lower stellar masses, from 0.3 dex at $log (M_*) \sim 9.4$ to 0.7 dex at $log (M_*) \sim 6.6$.}
  \label{fig:Ha_dist_Weisz}
\end{figure}

$L_{H\alpha}$ is strongly correlated with stellar mass, $M_*$, with the scatter around the mean increasing at lower stellar mass (Figure \ref{fig:Ha_dist_Weisz}). Thus, this distribution contains additional information about the SFH and we should be incorporating the $L_{H\alpha}$ distribution into our analyses of the SFH. Of course, the average of the distribution depends strongly on the stellar mass of the galaxy, and does not give any information about the relative changes in the SFR.  We therefore choose to subtract the trend of log$(L_{H\alpha})$, defining this as \del, plotted in Figure \ref{fig:Ha_dist_Weisz}. The value of \del\ tells us a galaxy's (nearly instantaneous) star formation rate, relative to the average, and the width of the distribution translates to a maximum burst amplitude, relative to the average. But it is particularly useful when examined {\it in conjunction} with the log(\huv), as log(\huv) gives us time information, indicating how quickly (or how recently) a given galaxy has increased or decreased its SFR.

\begin{figure}
\includegraphics[width=1\linewidth]{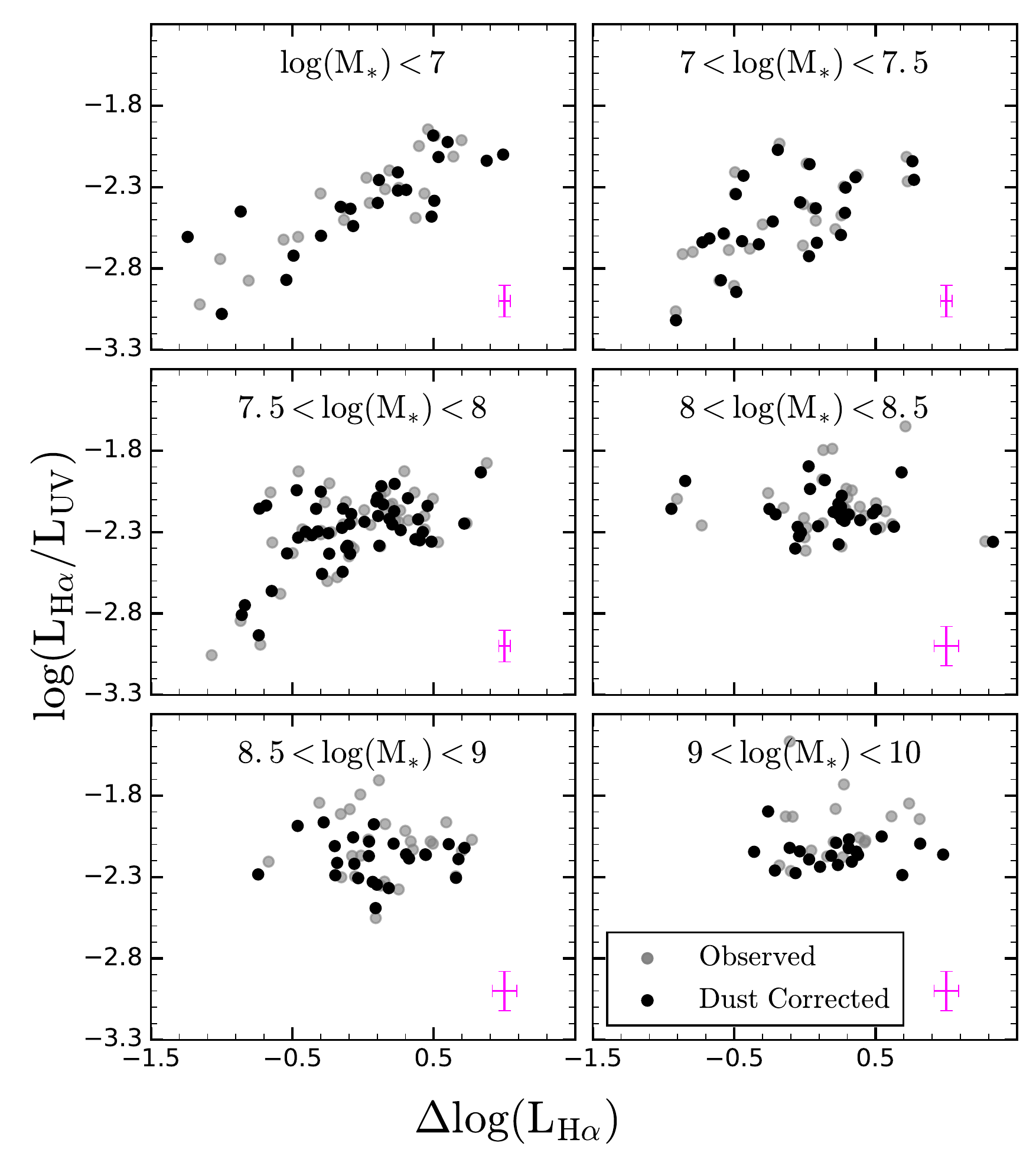}
\centering
\caption{log(\huv) vs. \del\ for the W12 sample in six different mass ranges. The grey and black points are the observed and dust corrected values respectively. Representative error bars of 10-20\% in $L_{H\alpha}$ and $L_{UV}$ derived from \citet{Lee_2009} is shown in the bottom-right of each subplot in magenta.} At high stellar mass, the galaxies span a narrow range in log(\huv). At low mass, the galaxies span a large range in log(\huv), and log(\huv) is positively correlated with \del. This correlations shows that, not only should these two observables be used to determine the properties of bursty star formation, but their 2-dimensional distribution should be considered. 
 \label{fig:deltaHa_Ha_to_UV_Weisz}
\end{figure}

In Figure \ref{fig:deltaHa_Ha_to_UV_Weisz}, we show the location of the W12 galaxies in the \del - log(\huv) plane in bins of stellar mass. The grey and black points correspond to the observed and dust corrected values, respectively. Note that the dust corrections are very small and do not change the shape of the correlations seen in Figure \ref{fig:deltaHa_Ha_to_UV_Weisz}. The representative errors of 10-20\% corresponding to $L_{H\alpha}$ and $L_{UV}$ are shown in the bottom-right of each subplot in magenta. These errors are derived from the typical dispersion in dust attenuation as a function of absolute B-band magnitude, as reported in \citet{Lee_2009}. Specifically, galaxies in our sample span $-18.4<M_B<-13$, for which the $L_{H\alpha}$ is uncertain by 10\% for stellar masses below $10^8 M_{\odot}$ and 20\% otherwise, and the $L_{UV}$ is uncertain by 20\% in the entire sample. As indicated before, the lower mass galaxies span a larger range in \del\ {\it and} log(\huv). But what is most informative is that the two measurements are highly correlated in galaxies with $\text{log}(M_*) < 8$. That is, we are not just fitting to the \del\ and log(\huv) 1-D distributions, but also considering the correlations between the two.   

We emphasize that we would like to use these distributions to infer SFHs of all of the galaxies in each mass bin.  In order to do that, it is important to be sure that all galaxies in the bin are behaving in a similar manner, and that these measurements represent random samples of the same SFHs.  One particular point of concern is the possibility that subsets of galaxies permanently reside above the SF main sequence (\del $> 0$) and others reside below the main sequence (\del$ < 0$), for example because of different average gas infall rates in different environments. If that were the case, there would be no difference in the log(\huv) ratio of galaxies above and below the main sequence. However, Figure \ref{fig:deltaHa_Ha_to_UV_Weisz} shows a strong correlation between log(\huv) and \del\ in low mass galaxies (log$(M_*) < 8$). Specifically, galaxies that have higher than average SFRs have log(\huv) $\sim-2.2$, indicative of constant or rising SFRs, and galaxies that have lower than average SFRs have log(\huv) $<-2.5$, indicative of declining SFRs. This indicates that all of the galaxies in a given stellar mass bin have similar star formation histories.

\subsection{The Exponential Burst Model}
\label{sec:exp}

The SF histories of galaxies are not periodic, top-hat bursts, and are instead stochastic in nature. Because such simple models do not represent the actual SFHs, it is difficult to ascertain whether the best-fit parameters of the model are physically meaningful. Therefore, we are not particularly interested in confining ourselves to a highly parameterized, unphysical star formation history. As discussed above, the amplitude of the bursts can be determined from the width of the $L_{H\alpha}$ distribution. What remains unknown are the timescales for the burst and quench phases. 

We have chosen to parameterize the rise and decline of a burst as an exponential in time such that

\begin{equation}
SFR(t)=
\begin{cases}
  e^{t/ \tau}, & \text{ if } 0 \leq t < D \\ 
  e^{-(t-2D)/ \tau}, &\text{ if }  D < t \leq  2 \times D
\end{cases}
\label{eq:2}
\end{equation}

\noindent where $\tau$ is the $e$-folding time, representing the typical timescale for significant change in the SFR. 
Part of our motivation for choosing an exponential burst model comes from the SFHs in some hydrodynamical simulations. For example star formation histories from the FIRE-2 simulations \citep{Hopkins_2018} indicate individual bursts with exponential rises and declines (see Figure \ref{fig:22_FIRE12b}).  For the sake of simplicity, we assume that the galaxies' SFR rise and decline are both described with the same $e$-folding time, $\tau$. With such a parametrization, we can target the timescale, independent of the absolute amplitude of the burst. Also, the SFHs are assumed to be periodic in repeating forms of Equation \ref{eq:2}.

\begin{figure*}
\includegraphics[width=1\linewidth]{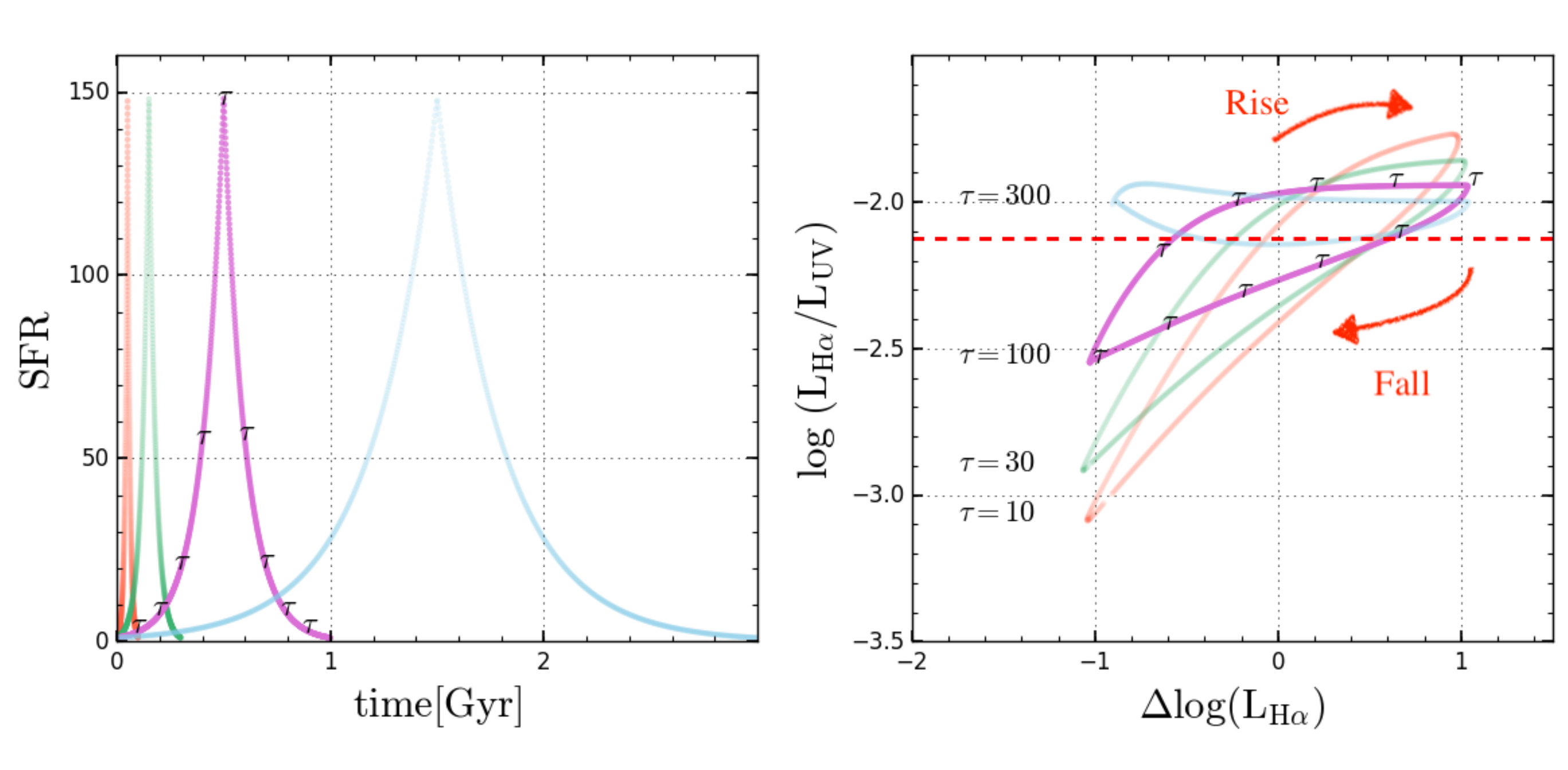}
\includegraphics[width=1\linewidth]{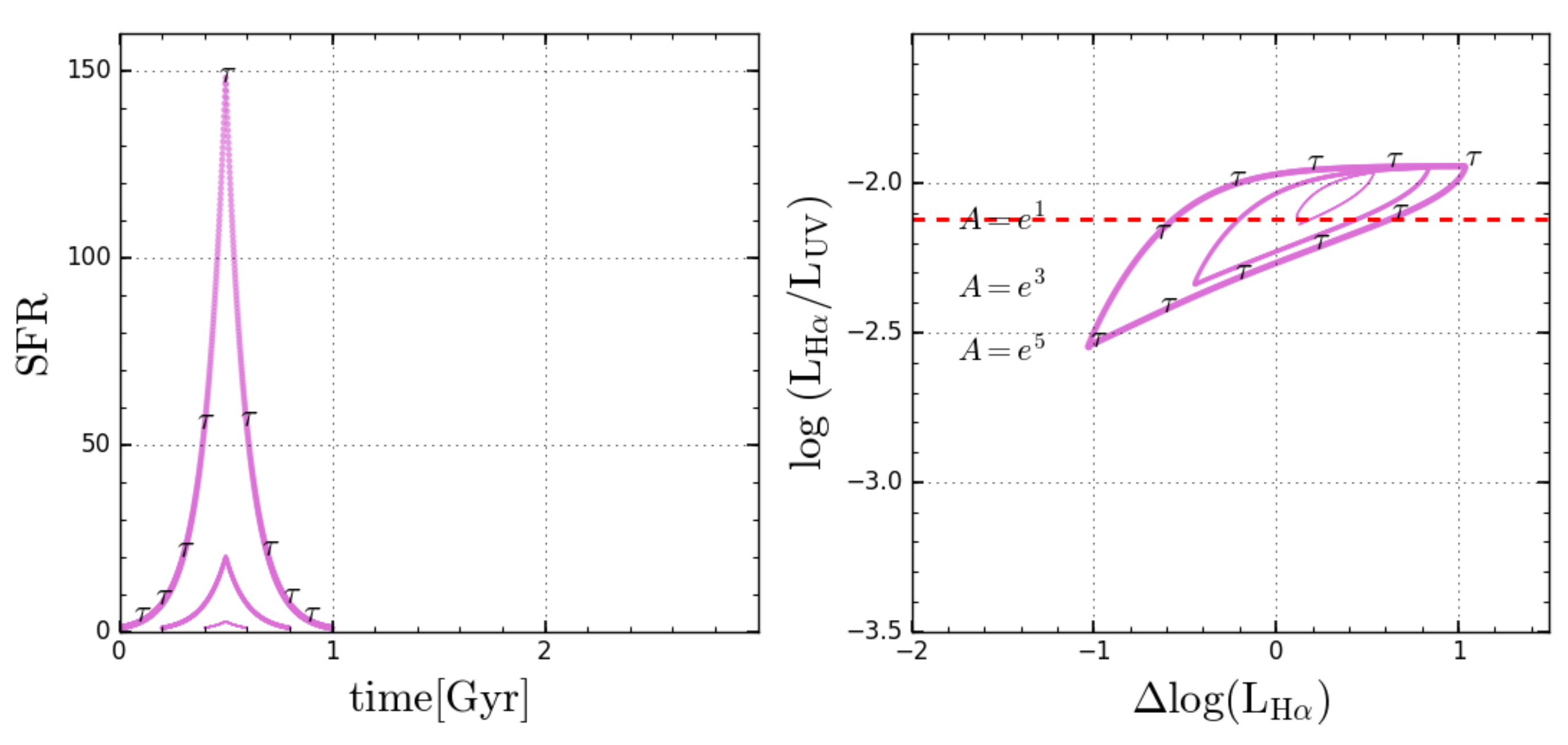}
\caption{Top Left: SF history of four exponentially rising and declining burst with varying $\tau=10, 30, 100, 300$ Myr and a fixed amplitude of $A = e^5=148$ (or equivalently a duration of $5\tau$). Top Right: The associated log(\huv) - \del\ for the SFHs in left panel. We highlight the $\tau=100$ Myr model, and place '$\tau$' symbols in intervals of $\Delta t = \tau$.  The equilibrium value (assuming a constant SFR) of log(\huv)=-2.12 is marked with red dashed line. As $\tau$ decreases, the log(\huv) - \del\ slope gets steeper. Bottom : Same as top panels except for models with fixed $\tau=100$ Myr and varying amplitudes of $A=e^1, e^3$ and $e^5$. The log(\huv) - \del\ slope is unchanged but the points span over smaller \del\ ranges.} Note that in an exponential model, the galaxies spend equal amounts of time in equal intervals of \del. 
\label{fig:20_6}
\end{figure*}

In the top left panel of Figure \ref{fig:20_6}, we plot model star formation histories with $\tau= 10, 30, 100, 300$ Myr and an amplitude of $A = e^5=148$ (or equivalently a duration of $5\tau$). In the top right panel, we plot the associated log(\huv) vs. \del. Here \del\ is calculated as the $log(L_{H\alpha})$ from \citet{Bruzual_Charlott2003} synthetic stellar population (BC03) model subtracted from the value of $log(L_{H\alpha})$ for a constant SFR that is equal to the {\it average} SFR of the exponential model. 

The $\tau=100$ Myr model is highlighted in purple to demonstrate the motion of a bursty galaxy in this observable space. We note that the observable ratios are dependent on previous star formation.  Thus, the plotted ratios are for a burst preceded by identical bursts. Obviously, as the galaxy SFR rises (burst phase), \del\ increases and the galaxy moves to the right in the plot.  As the galaxy SFR declines (quench phase), \del\ decreases and the galaxy moves to the left in the plot. The motion in log(\huv) is more complicated, depending on the timescale of change in SFR, $\tau$. For long timescales, $\tau > 300$ Myr, both of the luminosities in the log(\huv) ratio have time to react and accurately trace the SFR.  Therefore, log(\huv) does not significantly change as the SFR declines. However, if $\tau < 300$ Myr, $L_{UV}$ lags $L_{H\alpha}$ in tracing the decline in SFR, and log(\huv) will decrease as \del\ decreases. As $\tau$ decreases, the slope of the curves increase.  However, the slope saturates at $\sim 0.6$ for $\tau < 30$ Myr. Therefore, examination of galaxies in this observable space is only useful for identifying changes in SFR on timescales of $30<\tau<300$ Myr. The bottom panels of Figure \ref{fig:20_6} indicate the SFHs and associated log(\huv) - \del\ relation of three exponential burst models for the case where the e-folding time ($\tau$) is fixed to 100 Myr and the amplitudes are allowed to vary from $A=e^1, e^3$ and $e^5$. The model with larger amplitude ($e^5$) spans a larger \del\ axis, while the slope remains unchanged in all three models.

We note that in the exponential model, the galaxies are evenly spread in \del\ during the transition from quench to burst and vice versa. In contrast, the periodic top-hat burst model results in most galaxies having two \del\ values, with few points in between. The observed ratios are roughly evenly distributed in \del, similar to the exponential model.

\subsection{Results}
\label{sec:results}

\begin{figure}
\includegraphics[width=1\linewidth]{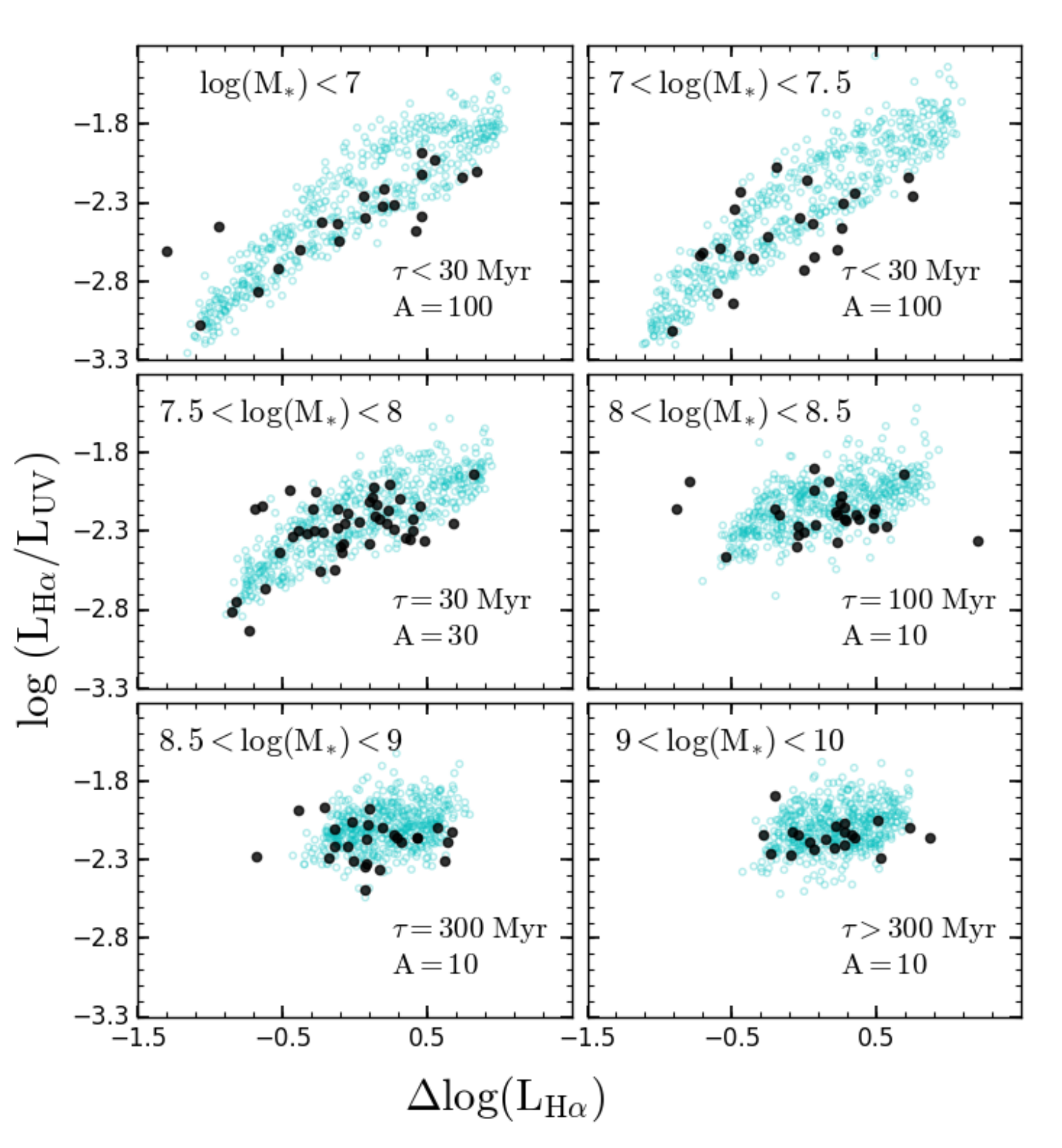}
\caption{Location of modeled galaxies in the log(\huv) - \del\ plane (cyan circles), undergoing periodic, exponential bursts of SF with different timescales, $\tau$, and amplitude, $A$, sorted by mass. Note that the time scales in the two lowest mass bins read $\tau<30$ Myr because the slope of the relation does not change below that time scale. We used $\tau=10$ Myr for these two mass bins. A 10-20\% systematic uncertainties corresponding to errors reported in Figure \ref{fig:deltaHa_Ha_to_UV_Weisz}} is added to each luminosity to mimic the observational uncertainties. The black points are the W12 observed data. The best-fit values of $\tau$ and A are written in the bottom right of each subplot. The time scale for changes in SFR increase with increasing stellar mass.
\label{fig:24_1}
\end{figure}

In Figure \ref{fig:24_1}, we again plot the W12 data in this log(\huv) - \del\ plane. As mentioned above, the data clearly show that even though the SFR of the high mass galaxies changes significantly, the log(\huv) ratio does not.  In the context of our exponential burst model, this indicates that the changes in SFR are slow, with $\tau \gtrsim 300$ Myr. At low mass, however, the log(\huv) is highly correlated with \del, suggesting that the SFRs change rapidly, with $\tau \lesssim 100$ Myr.  

In Figure \ref{fig:24_1}, we also plot the tracks (cyan) for values of $\tau$ and $A$ that best match the data in each mass range. The summary of these values are also reported in Table \ref{table:best_fit_params} for a better comparison. We note that we do not perform any probabilistic analysis to determine the best fit exponential models of each mass bin. Due to systematic uncertainties (e.g. extinction correction, metallicity, stellar models, escape fraction), the models may not precisely reproduce the data. We therefore determine (by eye) approximate model parameters that reproduce the slope and breadth of the data the observed log(\huv) vs. \del\ plane. A 10-20 \% observational uncertainty is added to the model (in both $L_{H\alpha}$ and $L_{UV}$) to mimic the true uncertainty in the data, as discussed in Section \ref{sec:lha}. The approximate values of $\tau$ and A are written in the bottom right of each subplot. Two trends are clear with mass.  First, the amplitudes -- the ratio of maximum to minimum SFR (or $L_{H\alpha}$) in galaxies of the same mass -- increase toward lower mass ($A\sim10$ for log$(M_*)>9$ to $A\sim100$ for log$(M_*)<7.5)$. Second, the timescales for changes in SFR ($\tau$) decrease with decreasing stellar mass, with $\tau>300$ Myr for log$(M_*)>8.5$ and $\tau<30$ Myr for log$(M_*)<7.5$. 

We note that, within the scatter, the slopes of the data in the log(\huv) - \del\ plane are roughly the same for all galaxies with log$(M_*)>8.5$ (slope $\sim0$) and for all galaxies with log$(M_*)<7.5$ (slope $\sim0.6$).
%These same slopes are shown in each subplot in Figure \ref{fig:24_1}.
As noted above, the log(\huv) ratio is most useful for identifying changes in SFR on timescales of $\tau \sim 100$ Myr. Thus, it appear that the stellar mass of log$(M_*)\sim8$ (at $z=0$) represents the critical mass at which the timescale for change in SFR is approximately $\tau \sim 100$ Myr. This is an observational result, and is true regardless of the physical mechanisms involved.

It is noteworthy that \citet{Meurer_2009} looked at the burstiness in the  log(\huv) - $\Sigma_{H\alpha}$ plane, which is very similar to the observables we analyze (replacing \del\ with $H\alpha$ surface brightness). In addition, they used Gaussian bursts and gasps, which are exponential in the beginning and end, similar to our model bursts. Indeed, their models show a similar behavior in log(\huv)-$\Sigma_{H\alpha}$ as our exponential models, especially in the gasp phase. However, their models have a very significant difference from ours - they assume a low-level, constant SFR when not in a burst or gasp. This assumption requires that the galaxies have a high log(\huv) before a new burst/gasp starts.  Ultimately, this assumption makes it impossible to reproduce the observed log(\huv) - $\Sigma_{R}$ (optical surface brightness) distribution of the galaxies, because it can not produce galaxies with both low log(\huv) (which can be produced only with a quick downturn in star formation) and low $\Sigma_{R}$ (which can not be produced with a quick downturn in star formation, if one assume that galaxies start with a high $\Sigma_{R}$). \cite{Meurer_2009} therefore concluded that the star formation history can not fully explain the low log(\huv) ratio in low surface brightness galaxies.  

Within the framework of our model, however, these low surface brightness galaxies with low log(\huv) are naturally explained by low mass galaxies \citep[which typically have low optical surface brightness,][]{de_Jong_2000} with rapidly declining SFR. Ultimately, we believe that our splitting of the sample into different mass bins (which, on average, have different optical surface brightness and different timescales), and our assumption that no baseline, constant SFR is required in our models, allows us to naturally explain the $L_{H\alpha}$ and $L_{UV}$ distributions with SFHs alone. 

There have also been investigations of the properties of burty galaxies at moderate-redshift ($0.4<z<1$) in CANDELS GOODS-N by looking at the $L_{H\beta}/L_{UV}$ \citep{2016_Guo}. In this paper, they observed a decreasing trend of $L_{H\beta}/L_{UV}$ towards low masses similar to the local sample of W12. They found a correlation between the $L_{H\beta}/L_{UV}$ and the $L_{H\beta}$-derived specific SFR (sSFR) of their sample, as evidence for bursty star formation.  They conculde that the galaxies in their sample are also bursty.  But the bursty SF occurs below a high mass threshold than at lower redshift ($M_*<10^9$ M$_{\odot}$ compared to $M_*<10^8$ M$_{\odot}$ at low redshift).

\begin{table}
  \centering
  \begin{tabular}{@{}ccc@{}}
    \toprule
    (1) & (2) & (3)  \\
    
    $log (M_{*})$ & $\tau$ [Myr] & $A$
    \\
    \midrule
    $\leq 7$ & $< 30$ & $100$\\
    $7-7.5$ & $< 30$ &  $100$\\
    $7.5-8$ & $30$ &  $30$\\
    $8-8.5$ & $100$ & $10$\\
    $8.5-9$ & $300$ & $10$\\
    $9-10$ & $> 300$ & $10$\\
      \bottomrule
  \end{tabular}
  \caption{Best fit values of the exponential burst parameters: the e-folding time ($\tau$) in column (2) and amplitude ($A$) in column (3) for six different mass bins from lowest to the highest stellar masses (column (1)).}
  \label{table:best_fit_params}
\end{table}

\section{Discussion}
\label{sec:discussion}

So far, we have constrained the parameters of bursty star formation by looking at the position of the observed galaxies in the log(\huv) - \del\ plane. We used an exponential SF model to reproduce the log(\huv) - \del\ distribution and found the best fit time scales and amplitudes of bursts as a function of stellar mass.

In this section, we compare our predictions with hydrodynamical simulations. We also consider the effect of stochastic IMF sampling within our log(\huv) - \del\ framework. Finally, we discuss the effect of the ionizing photon escape fraction and dust correction uncertainties.

\subsection{Comparison To Hydrodynamical Simulations}

Many high resolution hydrodynamical simulations of dwarf galaxies implement stellar feedback and supernovae, resulting in bursty SF \citet{Governato_2010, Dominguez_2015}. The effects of bursty SFHs on sample selection and interpretation of observables has been extensively investigated by \citet{Dominguez_2015} using the SFHs from hydro-dynamical simulations.
The FIRE-1 and FIRE-2 (Feedback In Realistic Environments) simulations \citep{Hopkins2014, Hopkins_2018}, implement prescriptions for a variety of stellar and supernovae feedback that are not tuned to reproduce observed scaling laws. Galaxies of different stellar masses are simulated, from dwarfs to Milky Way-like systems, reproducing the empirical relations between galactic observables such as the stellar mass-halo mass relation, the mass-metallicity relation, etc. Due to its high resolution and focus on stellar/supernovae feedback, the FIRE-2 simulations are well-suited for our goal of studying burstiness in dwarf galaxies.

\cite{Sparre_2017} studied $L_{H\alpha}$ and $L_{UV}$ as proxies of the SFR averaged over 10 and 200 Myr and found that the FIRE-1 simulations \citep{Hopkins2014} are more bursty than observed galaxies. Specifically, the FIRE-1 galaxies display a larger range in \huv\ than the observed galaxies of W12 between $8<\rm{log}(M_*[M_{\odot}])<9.5$. The FIRE-2 simulations \citep{Hopkins_2018} used improved numerical accuracy, resulting in more accurate treatments of cooling and recombination rates, gravitational force softening and numerical feedback coupling. However, the core physics is the same as FIRE-1. These enhancements lead FIRE-2 to produce more realistic bursts of SF. FIRE-2 also contains a larger sample of galaxies than the previous version, allowing a better understanding of variations in galaxy properties. The detailed properties of the individual simulations can be found in \citet{Hopkins_2018, El-Badry_2018}.

We use the star formation histories of the FIRE-2 galaxies, which span a similar range in stellar mass over their last 2 Gyrs (corresponding to $z=0.15-0$) to the W12 mass range, to determine their location and evolution in the log(\huv) vs. \del\ plane, and compare them with the observed W12 sample.

\begin{figure}
\includegraphics[width=1\linewidth]{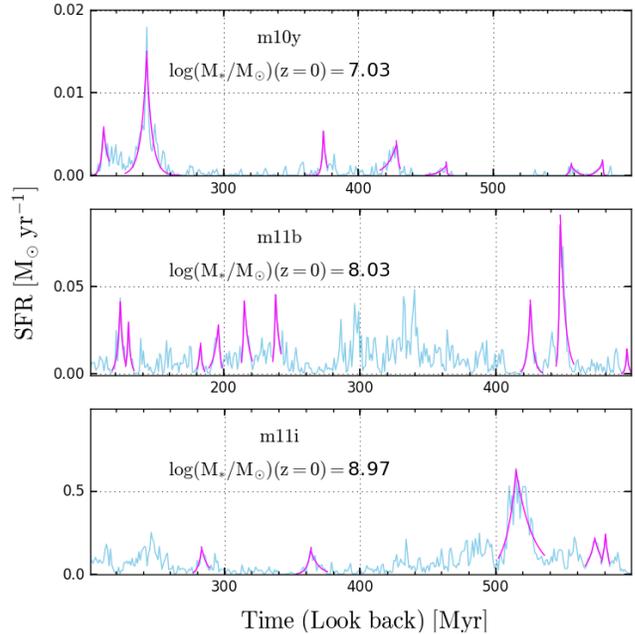}

\caption{Star formation histories of three galaxies in the FIRE-2 simulations \citep{Hopkins_2018} (cyan). The name and stellar mass of the galaxies at $z=0$ are labeled in the top-left of each panel. The magenta lines are the best fit exponential functions to the individual bursts. The marked bursts all have e-folding time scale of $\tau$ below 15 Myr.}
\label{fig:22_FIRE12b}
\end{figure}

Figure \ref{fig:22_FIRE12b} shows three examples of star formation histories of FIRE-2 simulated galaxies, whose name and stellar mass at $z=0$ are written on the top-left of each panel. We specifically chose galaxies with a different range of stellar masses. The galaxies have bursts of SF with a short amount of time in the peak. We see that galaxies rise and decline quickly in a roughly exponential form. For illustration purposes, we marked some of these exponential bursts in Figure \ref{fig:22_FIRE12b} in magenta and fit equation \ref{eq:2} to individual bursts (using a curve fit Scipy package from Python) in order to find an estimate of rising and falling timescales for these galaxies. Based on this simple fitting, all three of these galaxies indicate time scales of $\tau < 15$ Myr.

The $L_{UV}$ and $L_{H\alpha}$ fluxes are derived by convolving the simulated star formation histories over the last 2 Gyrs with the H$\alpha$ and UV luminosity evolution from single stellar population models \citep{Bruzual_Charlott2003} with stellar metallicity of $0.2Z_{\odot}$. This is the same method as described for our parametrized bursty star formation in the Appendix. We note that in the low mass galaxies in which we are most interested, the vast majority of stars are formed in-situ and not acquired in mergers \citep{Fitts_2018}. Furthermore, most of the mergers happened at $z>3$, so the accreted stars are old and will not affect the $L_{H\alpha}$ or $L_{UV}$ \citep{Fitts_2018} calculations at late times.

\begin{figure}
\includegraphics[width=1\linewidth]{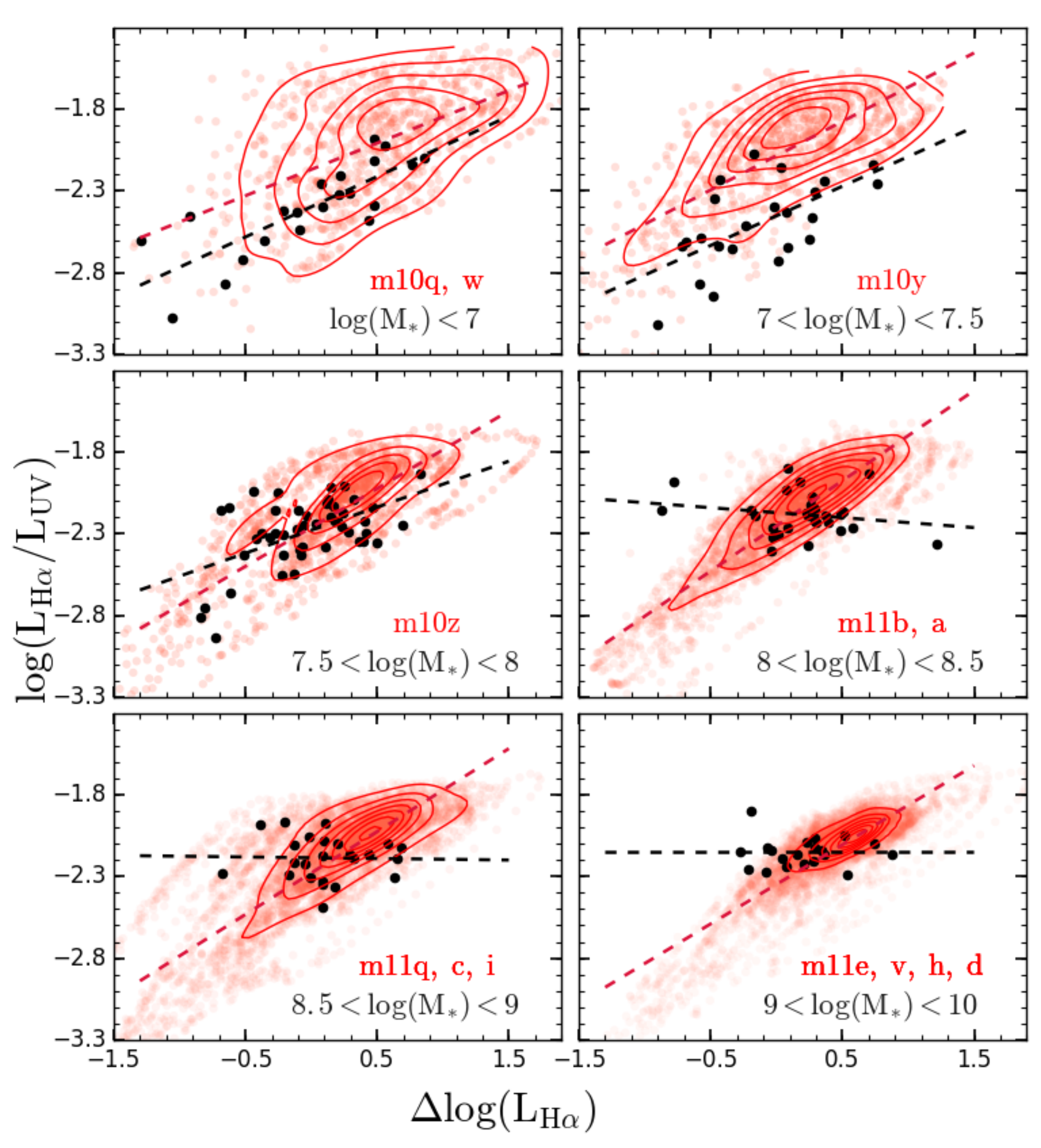}
\caption{The log(\huv) vs. \del\ relation for FIRE2-simulated galaxies (red circles and contours) and W12 data for local galaxies (black circles). The FIRE-2 galaxies are selected to be within the specified mass range at z=0. The names of the FIRE-2 galaxies are labeled in red. Red and black dashed lines are the best-fit line to the simulated and observed galaxies, respectively. We note that in the lowest mass galaxies (log$(M_*)<7)$, a significant fraction of galaxies exhibit no star formation for long periods of time and would lie off of the bottom left of the plot. The simulated galaxies all span a similar range in \del\ as the observed galaxies, suggesting similar burst amplitudes. However, the simulated galaxies at higher mass show a positive correlation between log(\huv) and \del that is not seen in the observed galaxies, suggesting that the timescales for change in SFR is too short in simulated galaxies.This is also evident from the slope of the red dashed lines in these mass bins.}
\label{fig:5_1}
\end{figure}

In figure \ref{fig:5_1} we make the log(\huv) vs. \del\ plot for FIRE2-simulated galaxies, where the color indicates the density of points at any location on the plot. Overlaid is the W12 observed sample in black circles. We calculate the median $L_{H\alpha}$ from the linear fit to the $L_{H\alpha}$ vs. mass relation for FIRE-2 galaxies and subtract this to determine \del. The names of the plotted FIRE-2 galaxies are labeled in red.  One common feature that can be seen in both the observations and simulation is that the low mass galaxies exhibit a larger spread in the log(\huv) vs. \del\ space while the spread becomes smaller at higher masses. Nonetheless, there are some discrepancies between them.

As mentioned before, the observations show no relation between \huv\ and \del\ above $10^8$ M$_{\odot}$, but show a strong correlation below $10^8$ M$_{\odot}$. This is seen in the slope of the best-fit lines to the data in Figure \ref{fig:5_1}. However, the simulated galaxies show a strong correlation, with a similar slope, at all stellar masses (red dashed line in Figure \ref{fig:5_1}. Comparison with our exponential burst models (Section \ref{sec:exp}) on the FIRE-2 galaxies, indicates that the bursts/quenches have e-folding times of $\tau$ below 30 Myr in the simulated galaxies at all masses. However, the W12 data suggest that the e-folding time increases toward higher stellar masses. In particular, at stellar masses above $10^8$, the FIRE-2 simulations have far faster bursts ($\tau < 30$ Myr) than the timescales implied by the observation of real galaxies ($\tau \gtrsim 300$ Myr). We therefore suggest that the rapid bursts in the FIRE-2 simulations of more massive galaxies should be examined to determine possible shortcomings in the existing feedback prescriptions. 

We note here that although the log(\huv) and \del\ distributions are similar to the observed galaxies, it is the examination of the log(\huv) - \del\ 2-dimensional space that allows us to recognize that the time scales for the bursts in more massive galaxies may not be correct. 

One other difference between the simulation and observation that becomes more significant at masses below $10^{7.5} M_{\odot}$ is that the simulation produces larger log(\huv) than what is observed in real galaxies. We address this offset in section \ref{sec:f_escape}.

\begin{figure*}
\includegraphics[width=1\linewidth]{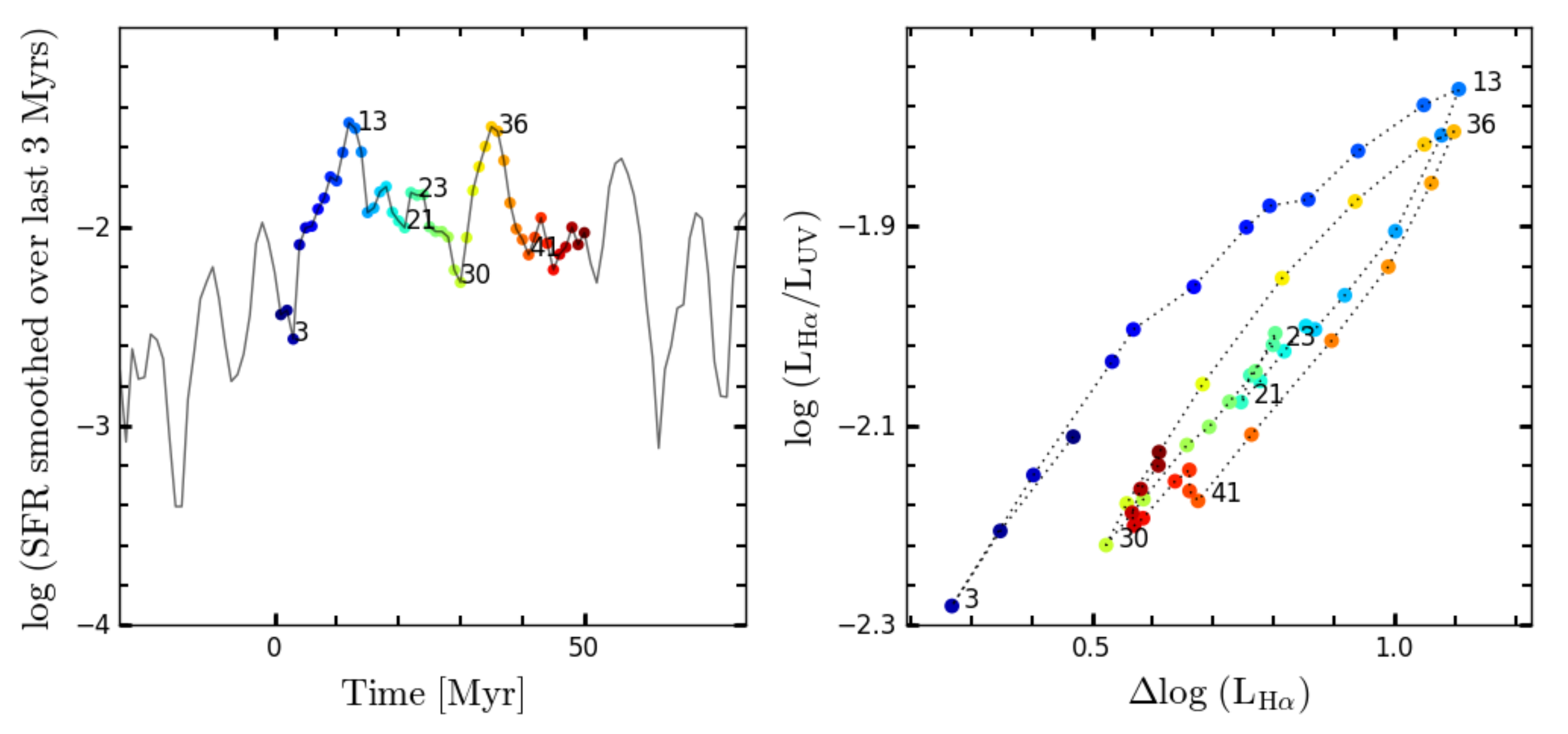}

\caption{ Left: The SFH smoothed over the last three Myrs for a 50 Myr segment of the recent SFH in galaxy ``m11b" with $10^{8}M_{\odot}$ stellar mass at present ($z=0$). Right: The log(\huv) vs. \del\ associated with the original (un-smoothed) SFH on the left panel. Colors and numbers represent the same star formation time on both panels. This is clear that the log(\huv) - \del\ traces the SFR change when SFR is smoothed over the last three Myrs. As star formation rises  and declines, it moves diagonally from bottom-left to top-right and back in log(\huv) - \del\ plane respectively.}
\label{fig:Fire_sfh_huv_del}
\end{figure*}

We choose galaxy ``m11b," with stellar mass of $10^{8} M_{\odot}$ at $z= 0$, as an example to demonstrate the effect of star formation rate change on the log(\huv) - \del\ relation. In Figure \ref{fig:Fire_sfh_huv_del}, we plot a 50 Myr segment of the recent SFH of this galaxy and marked each Myr in a color gradient as the time advances from dark blue to dark red as is shown on the left plot. The SFH at each point is averaged over its last three Myrs as it takes about three Myrs for the $L_{H\alpha}$ to react to the change in star formation rate. On the right panel, we plot the log(\huv) - \del\ associated with the original (un-smoothed) SFH for which the colors represent the corresponding time on the left panel of the same color. When the SFR begins to rise, the galaxy moves in the log(\huv) - \del\ plane from bottom-left to the top-right (3 to 13, 21 to 23, 30 to 36 Myrs). When the SFR declines quickly, the galaxy moves back toward the bottom left (13 to 21, 23 to 30, 36 to 41 Myr). Moreover, the larger amplitude changes in SFR give rise to larger changes in the log(\huv) - \del\ plane (compare the large amplitude decline from 36 to 41 Myr to the smaller amplitude decline from 23 to 30 Myr). This behavior is consistent with the predictions from our model.

\subsection{Stochastic IMF Sampling}
So far we have discussed the luminosity distributions from the perspective of bursty star formation, assuming that the IMF is well-sampled in all galaxies at all SFRs. However, it may be the case that stars form in a stochastic manner such that low mass galaxies with a limited amount of gas form massive stars less frequently. Thus, it may not be appropriate to assume a fully sampled mass function in low mass galaxies with low SFRs. This will result in a deficit of very high mass stars, and can significantly change the log(\huv) and \del distributions, even in a galaxy with a constant SFR. If this effect is large, it can incorrectly cause us to assume that a galaxy is undergoing bursty star formation. 

%Based on some studies \citep{Pflamm-Altenburg2007} arguing that stochastic IMF sampling can explain the observed turn-down in the log(\huv) vs. SFR for galaxies with $log$ $(SFR[M_{\odot} \text{yr}^{-1}])<-1.5$. This SFR is equivalent to $log(L_{H\alpha})<39.7\ [erg/sec]$, which corresponds to galaxies of masses below $\sim10^8 M_{\odot}$ in the W12 sample. 

Stars are born in star clusters and the distribution of clusters is called the initial cluster mass function (ICMF). The ICMF is typically modeled as a power-law, and observations suggest a slope of 2 \citep[][]{Zhang_1999, Fall_2009, Lada_2003}. Clusters of different masses are formed according to the probability given by the ICMF. Then, stars in each cluster are formed according to the probability given by the initial stellar mass function (ISMF).

Because some studies have indicated that IMF sampling can explain some of the observed log(\huv) distributions, we attempt here to model stochastic IMF sampling in the W12 galaxies. SLUG \citep[Stochastically Lighting Up Galaxies][]{daSilva_2012,daSilva_2014} is a code that considers stochastic sampling of both the ICMF and, then, the ISMF to determine the spectrum of star-forming galaxies. We use SLUG to determine the effects of ICMF and ISMF stochastic sampling to determine if poor sampling of the high mass ends is responsible for the large scatter in log(\huv) that we observe in low mass galaxies. We use stellar metallicities from Table \ref{table-1}, assume a power-law ICMF with slope of $\beta=2$, and Padova stellar tracks with thermally pulsing AGB stars and a Chabrier ISMF. In each mass bin, we use the median $L_{H\alpha}$ of all galaxies to determine a typical SFR in that mass bin, assuming the \citet{Kennicutt_1998a} conversion. The SLUG code takes the input (constant) SFR, and stochastically produces stars, outputting spectral energy distributions every 5 Myr. We assume that all ionizing photons result in a photoionization, and calculate the resulting $L_{H\alpha}$ based on case-B recombination.  We determine the \del\ distribution by subtracting the median from the $L_{H\alpha}$ distribution.

 \begin{figure}
\includegraphics[width=1\linewidth]{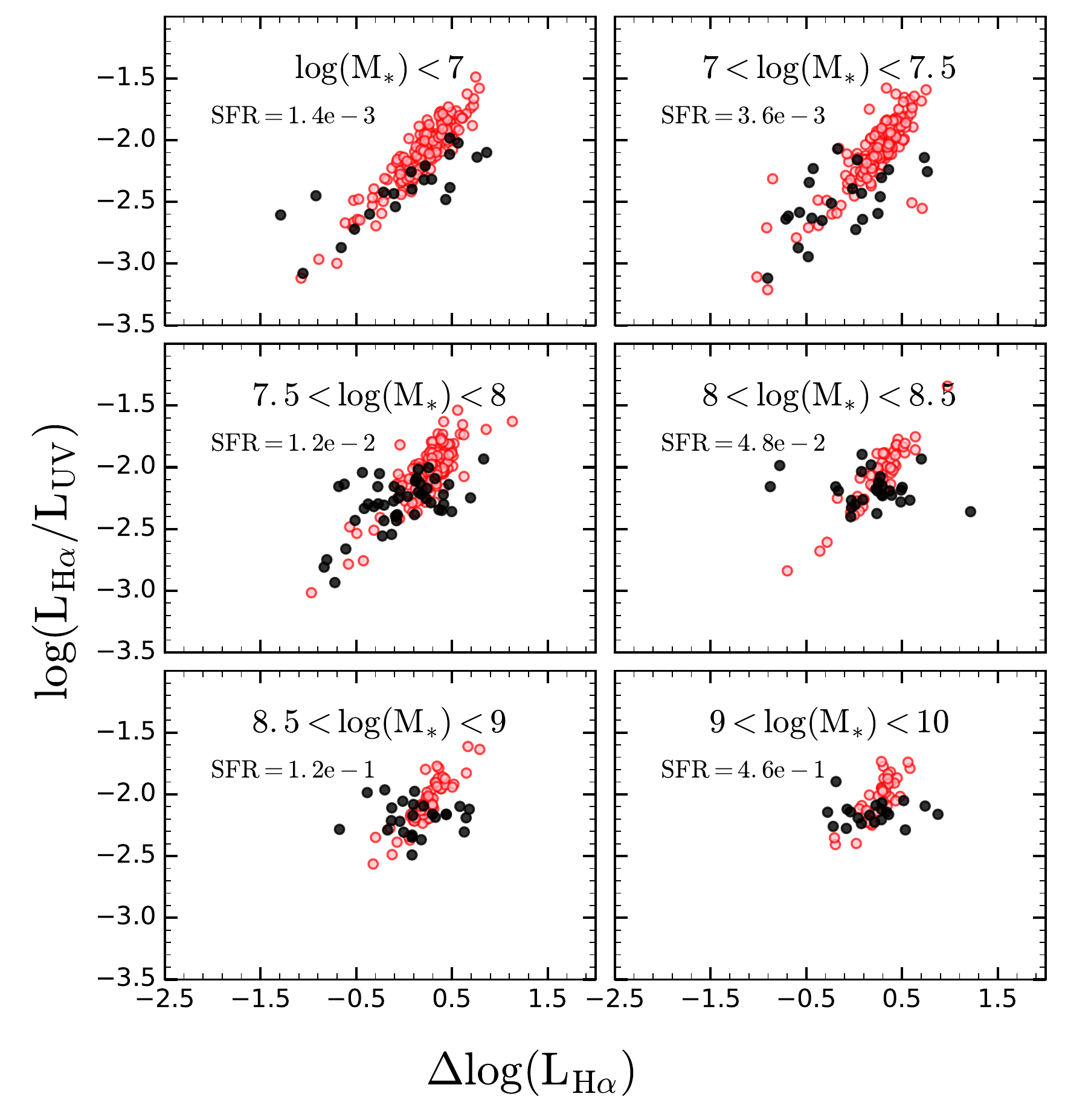}
\caption{The log(\huv) vs. \del\ relation due to stochastic sampling of the ICMF and ISMF. Red points are SLUG-synthesized galaxies and black points are the W12 observed galaxies. The input (constant) SFR is derived from the average $L_{H\alpha}$ of the W12 sample for each mass bin. Though the SFR is constant, the luminosities change considerably as high mass stars are stochastically created. When high mass stars are abundant, the galaxy is high in both log(\huv) and \del.  When high mass stars are not abundant, the galaxy is low in both log(\huv) and \del. Thus, there is a tight correlation between log(\huv) and \del, with a slope of $\sim1$ in all mass bins.}
\label{fig:SLUG}
\end{figure}
 
In Figure \ref{fig:SLUG}, red circles are the output SLUG data and black circles are the W12 observed data. In this figure, we show the evolution of galaxies with constant SFR in the log(\huv) vs. \del\ plane due to stochastic formation of massive stars. When high mass stars are abundant, the galaxy is high in both log(\huv) and \del. When high mass stars are not abundant, the galaxy is low in both log(\huv) and \del. Thus, there is a tight correlation between log(\huv) and \del, with a slope of $\sim1$ in all mass bins. We note that a slope of $\sim1$ suggests that $L_{H\alpha}$ is changing significantly, while the $L_{UV}$ is roughly constant. 

The SLUG points roughly produce the width of the observed \del\ distribution. However, SLUG significantly overpredicts the observed slope of the log(\huv) vs. \del trend (slope of 1 vs. $\sim0.5$). Therefore, it appears that bursty star formation is responsible for most of the observed spread in the log(\huv) and \del\ distributions. This is somewhat surprising, as the very low SFRs in our low mass bins should suggest incomplete sampling of the high mass end of the ISMF.  This may be reconciled if star-forming clusters in gas-rich, low mass galaxies are ``top-heavy'' relative to the assumed ICMF above. 

\citet{Eldridge_2012} added the effect of binary stars to the stochastic IMF sampling and looked at the log(\huv) distribution of the W12 sample. Binary star mergers and mass transfer both produce more massive stars than were present in the initial population, which blurs some of the observational differences between different IMF sampling in their models. However, they conclude that the scatter in the log(\huv) distribution is not due to the IMF sampling method but it depends more on the bursty star formation history of each individual galaxy.

Furthermore, more constraints can be imposed on IMF sampling by limiting the ISMF and ICMF at the high mass end, such that the maximum stellar mass in star clusters is limited by the mass of the cluster and that the mass of the cluster itself is constrained by the SFR. This method is referred to as the integrated galactic IMF \citep[IGIMF][]{Kroupa_Weidner_2003}. \citet{Pflamm-Altenburg2007, Pflamm-Altenburg2009, Lee_2009} argue that the IGIMF  explains the observed log(\huv) distribution of their sample. However, \citet{Fumagalli2011} suggest that the imposed high mass limits of both the ICMF and the ISMF in the IGIMF models lead to a dramatic reduction in the luminosity scatter at low SFRs, which is inconsistent with observations \citep{Andrews_IMF_2013, Andrews_IMF_2014}. 

\subsection{Escape of Ionizing Photons}
\label{sec:f_escape}
We note that our models consistently extend to higher log(\huv) values than are observed at all stellar masses (see Figure \ref{fig:24_1}), but especially at log$(M_*)<8$. This ratio is affected by stellar metallicity, so part of this discrepancy could be alleviated if we assumed higher metallicity values. However, we are using metallicities that are consistent with measured mass-metallicity trends in low redshift galaxies \citep{Andrews_2013} and, in any case, the offset between the observed galaxies and models in the low-mass galaxies ($\sim 0.3$ dex) is larger than can be explained by stellar metallicity alone \citep[$\sim0.1$ dex difference between 0.2 Z$_{\odot}$ and Z$_{\odot}$,][]{Bruzual_Charlott2003}. One possible explanation is that a significant fraction of the ionizing photons (roughly half) is escaping from low-mass galaxies in the peak of their burst phase, resulting in lower observed $L_{H\alpha}$.  Such a high escape fraction has only been observed in a handful of galaxies with extreme star formation surface densities and specific star formation rates \citep{Izotov_2018, Vanzella_2016, Vanzella_2018}, unlike the more typical galaxies in our sample. However, we note that the escape fraction has not been probed in galaxies with such low star formation rates ($SFR < 0.1$ M$_{\odot}$ yr$^{-1}$), as the Lyman continuum flux would be difficult to detect with current instrumentation. There has been an indirect search for escaping ionizing photons (via a deep search for faint H$\alpha$ recombination in the nearby circumgalactic medium of three nearby dwarf galaxies. But there was only 5\% more $L_{H\alpha}$ emission identified \citep{Lee_2016}.  This still leaves open the possibility that the ionizing photons may escape to a much larger radius ($> 2$ kpc). The possibility that typical low mass galaxies have high ionizing photon escape fractions is intriguing and has important implications for reionization and the subsequent evolution of the ionizing background. 

Another possible explanation is that the high mass end of the IMF is not being fully sampled, creating a deficit of the most massive stars and suppressing $L_{H\alpha}$. However, in Figure \ref{fig:SLUG}, we can see that modeling of the stochastic sampling of the IMF still produces a large number of galaxies above the observed distribution of log(\huv) at all masses. Therefore, it appears that IMF sampling can not fully explain the discrepancy.

\subsection{Dust Attenuation}
 We caution that the log(\huv) measurements contain potential systematic uncertainties arising from the application of a uniform dust attenuation curve.  We know that even galaxies with similar masses and star formation histories show considerable scatter in their extinction curves \citep{Meurer_1999}. This will increase the scatter in the log(\huv) vs. \del\ plane, but does not significantly affect our interpretation, as it ultimately depends on the slope of the distribution in that plane. In addition, there is some evidence that attenuation curves may change as a function of stellar mass and/or SFR \citep{Reddy_2015}. If the variations in attenuation curves are a function of stellar mass, this will not affect our results, as we have binned our data in narrow stellar mass ranges.  However, if the attenuation curves change as a function of SFR, then this would change the relative attenuation of $L_{UV}$ and $L_{H\alpha}$ {\it within a stellar mass bin}, which could change the slope of the distribution in the log(\huv) vs. \del\ plane.

\section{Summary}

In this paper we studied the phenomenon of bursty star formation in low-redshift dwarf galaxies and attempted to determine the parameters (amplitudes and timescales) of the bursts. The parameters can be measured by comparing star formation rate indicators (e.g., $L_{H\alpha}$ and $L_{UV}$) that are sensitive to different timescales. For our analyses, we used the data from \citet{Weisz12}, which includes extinction-corrected $L_{H\alpha}$ and $L_{UV}$ measurements for local galaxies with a large range in stellar mass ($10^{6}<M_{*}<10^{10} M_{\odot}$). 

First, we fit to the same top-hat periodic burst model of \citet{Weisz12} to determine the period, duration, and amplitude of the bursts. We improved the analysis by 1) using more appropriate sub-solar stellar metallicities, 2) expanding the probed parameter space and 3) using a likelihood analysis to better determine parameter uncertainties and degeneracies. We found that the results were broadly similar to those of \citet{Weisz12}, but with significantly longer durations and periods.  Moreover, we found that the parameters had significant uncertainties and degeneracies, with period and duration being highly degenerate. We therefore argued that it is not sufficient to use only a single log(\huv) distribution to constrain a three-parameter burst model.

We showed that a galaxy's location on the star-forming main sequence is correlated with the log(\huv) distribution, strongly suggesting that all of the galaxies are exhibiting similar star formation histories. Thus, we argue that the \del\ distribution (i.e., the log$(L_{H\alpha})$ deviation from the mean log$(L_{H\alpha})$ should be used to estimate burst parameters, in addition to the log(\huv) distribution. Indeed, the two parameters are correlated, and the motion of galaxies in the 2-dimensional log(\huv)-\del\ plane gives significant insight into the timescale for variations in the star formation rate. 

In order to avoid highly parameterized star formation histories, we look for a model with the least numbers of parameters that informs us most about the physical characteristics of the bursts. So we instead compare to exponential bursts with two parameters: 1) the time scale of rising and falling of the SFR ($e$-folding time, $\tau$) and the maximum amplitude of the bursts ($A$). We find that galaxies with stellar masses less than $10^{7.5} M_{\odot}$ undergo large and rapid changes in SFR with timescales of $\tau < 30$ Myr and maximum amplitudes of $A \sim 100$ while galaxies more massive than $10^{8.5}$ $M_{\odot}$ experience smaller, slower changes in SFR with $\tau >300$ Myr and $A\sim 10$.
 
We also calculated the log(\huv)-\del\ relation for galaxies in the FIRE-2 hydrodynamical simulations and found that these galaxies exhibit short timescale ($\tau < 30 Myr$) changes in SF at all mass ranges. Though the amplitudes of these bursts agree well with the observed \del distributions, such short bursts are different from the long timescales of $\tau > 100 Myr$ that we inferred for galaxies above $10^8 M_{\odot}$. Future improvements to the simulations should look carefully at what is causing such short bursts in the more massive galaxies ($M_* \gtrsim 10^8 M_{\odot}$).

Furthermore, we examined the stochastic IMF sampling models using the SLUG code \citep{daSilva_2014} and found that the simulated log(\huv)-\del distributions were significantly steeper than the observed distributions. Therefore, stochastic sampling of the IMF may help explain some of the scatter in log(\huv)-\del, but the assumptions of the mass function of the clusters or stars may need to be revised.

Finally, we note that measurements of $L_{UV}$ already exist for large samples of galaxies from $z\sim1$ up to the earliest epochs ($z\sim8$). Soon, with the advent of JWST, WFIRST and the next generation of 30-meter-class ground-based telescopes, measurements of $L_{H\alpha}$ will be routine, even for dwarf galaxies at high redshift. As galaxies at early epochs will have very different conditions (smaller physical sizes, higher gas fractions, more metal-poor stars), it will be necessary to use a similar analysis to determine the properties of bursty star formation at high redshift. This will in turn allow us to better interpret these observables to more accurately determine physical properties.

\section{Acknowledgements}

 We thank the anonymous referee for providing useful comments that helped improve the quality of this paper. NE and BS acknowledge support from Program number 13905 provided by NASA through a grant from the Space Telescope Science Institute, which is operated by the Association of Universities for Research in Astronomy, Incorporated, under NASA contract NAS5-26555. DRW is supported by an Alfred P. Sloan Fellowship and acknowledges the Alexander von Humboldt Foundation. 

\appendix
\label{sec:appendix}

\section{Details of determining SF model parameters based on log(\huv) distribution}

We first describe the \citet{Weisz12} method of constraining burst parameters, and then explain our improvements to the method. 

\citet{Weisz12} divided their observed sample into five stellar mass bins from $~10^{6}$ to $~10^{11} M_{\odot}$ and assumed that all galaxies of similar stellar masses have star formation histories with the same parameters. They defined their SF models as top-hat, periodic bursts superimposed on a constant, baseline SFR and assumed stellar metallicity of $1Z_{\odot}$. They then determined the log(\huv) distribution from models with different burst periods (P), durations (D), and amplitudes (A) and used a Kolmogorov-Smirnov test to identify the parameters that best reproduce the observed log(\huv) distribution in each mass bin. W12 finely sampled the parameter space, resulting in 1466 SFH models. The main conclusion of W12 is that galaxies with the lowest stellar masses have higher amplitude bursts (A $\sim 30\times$ the baseline rate), relatively long durations (D $\sim 30-40$ Myr), and long periods ($P = 250$ Myr). The highest mass bins are characterized by almost constant SFRs with an occasional modest burst favoring SF models with short duration (D $\sim$ 6 Myr) and modest amplitudes (A $\sim 10$).

We use the same periodic, top-hat burst model as \citet{Weisz12}, with the same parameters (P, D, A). The duration is restricted to be less than the period. We note that the value of the baseline SFR is irrelevant, as the relevant factor affecting luminosity {\it ratios} is the ratio of the amplitude SFR to the baseline SFR. From observations, we have a sample of log(\huv) for 185 local galaxies ranging from $5 < log(M_* [M_{\odot}])<11 $ in mass. The observed sample is classified into six mass bins.

For each individual set of model parameters, we build a sample of 10,000 log(\huv) values for that SFH, selected in $5 Myr$ time intervals. This is a way of representing an ensemble of galaxies in the real world that are caught during different stages of the star formation history assuming that they all share the same SFH. We do not sample the first 200 Myrs of star formation, to ensure that the $L_{UV}$ is not biased low as it would not have enough time to build up a full sample of A-stars. To do this, we first calculate the spectral energy distribution (SED) as a function of time for a single stellar population \citep{Bruzual_Charlott2003} using a Chabrier IMF \citep{Chabrier_2003}, Padova isochrones \citep{Bertelli_1994, Bressan_1993, Fagotto_1994}, and a constant stellar metallicity. For each output SED, we calculate $L_{UV}$ as the average of the continuum between 1460-1540 \AA\ and $L_{H\alpha}$ by determining the Hydrogen-ionizing photon production rate and assuming case B recombination. We then convolve our $L_{UV}(t)$ and $L_{H\alpha}(t)$ curves with the $SFR(t)$ curves to determine the {\it intrinsic} (unobscured by dust) log(\huv) as a function of time.

To take into account the observational uncertainties, we add 10-20\% Gaussian errors to $L_{UV}$ and $L_{H\alpha}$ due to statistical uncertainties arising from dust attenuation corrections (See section \ref{sec:lha}.)

Finally, we determine the probability of obtaining our measured distribution of log(\huv) given the modeled distribution for the set of parameters (D, P, A). We then determine the set of parameters that maximizes the probability space and marginalize them to determine the 1-D posteriors distributions for each individual parameter.

\begin{center}
\begin{deluxetable}{c c c}
\tablecaption{Stellar metallicities used in models.}
\tablehead{
\colhead{Stellar Mass} &  \colhead{Measured $Z$\tablenotemark{a}}   & \colhead{Model $Z $\tablenotemark{b}} \\
\colhead{$log(\frac{M}{M_{\odot}})$} & \colhead{[$Z_{\odot}$]}  & \colhead{[$Z_{\odot}$]}
}
\startdata
 $\leq 7$ & N/A & $0.2 $ \\
 $7-7.5$ & $N/A- 0.10$ & $0.2 $ \\
 $7.5-8$ & $0.10- 0.17$ & $0.4 $ \\
 $8-8.5$ & $0.17- 0.31$ & $0.4 $ \\
 $8.5-9$ & $0.31- 0.46$ & $0.4 $ \\
 $9-10$ & $0.46- 0.70$ & $0.4 $ 
\enddata
\label{table-1}
\tablenotetext{a}{Measured gas-phase metallicity for each stellar mass bin in \citet{Andrews_2013}}
\tablenotetext{b}{Closest stellar metallicity model in \citet{Bruzual_Charlott2003}. Note that we use $0.4 Z_{\odot}$ for log($M_*)=7.5-8$ (instead of 0.2 $Z_{\odot}$) in order to fit the observed log(\huv) distribution.}
\end{deluxetable}
\end{center}

In order to select the appropriate metallicity for each stellar mass bin, we refer to the mass-metallicity relation from \cite{Andrews_2013}, assuming solar Oxygen abundance to be $8.86$. In Table \ref{table-1}, the approximate metallicity of each mass bin is determined except for the lowest mass bin due to lack of data points in this mass range. We choose to use the metallicities from the existing libraries in \citet{Bruzual_Charlott2003} that are closest to the values in column 2 of Table \ref{table-1}. The chosen metallicities are given in column 3. For $M\leq 10^{7}\ M_\odot$ we assumed 0.2 solar metallicity because the next smallest metallicity available in the BC03 is 0.02 $Z_{\odot}$, far lower than we expect for galaxies at those stellar masses. The results of this model is discussed in sec. 3.3.

\subsection{Improvements On Fits To Periodic SF History}
\label{w12_improvements}

As discussed above, the W12 study made significant progress in fitting to the 
$log(\frac{L_{H\alpha}}{L_{UV}})$ distribution to determine SF burst model parameters.  However, we have identified ways in which we can enhance the W12 analysis.

First, the stellar metallicity was assumed to be solar for all mass ranges. This can change the estimated SF model parameters because stellar metallicity affects the surface temperature of stars such that more metal-rich stars are cooler at a given mass. This therefore affects the amount of ionizing to non-ionizing photons. To address that, we explore the effects of using more realistic (lower) stellar metallicities on parameter estimation for different mass ranges based on the metallicities estimated in \citet{Andrews_2013} (see Table \ref{table-1}).

Second, in order to explore the model parameter space, 
W12 used a two-sided K-S (Kolmogrov-Schmirnov) test. However, we use likelihood approach that allows us to better determine parameter uncertainties and degeneracies.
 
 Third, the best-fit parameters from W12 were at the edge of the explored parameter space, so it is possible that the actual best-fit parameters are out of the explored range. Therefore, we expanded the parameter space (by a factor of five in duration, period, and amplitude) in order to guarantee that we explore all parameter space far enough that encompasses the best-fit parameter.

\subsection{Results of Best Fit Burst Parameters}
\label{Appendix_2}
\begin{figure*}
 \begin{center}
					
\includegraphics[width=1\linewidth]{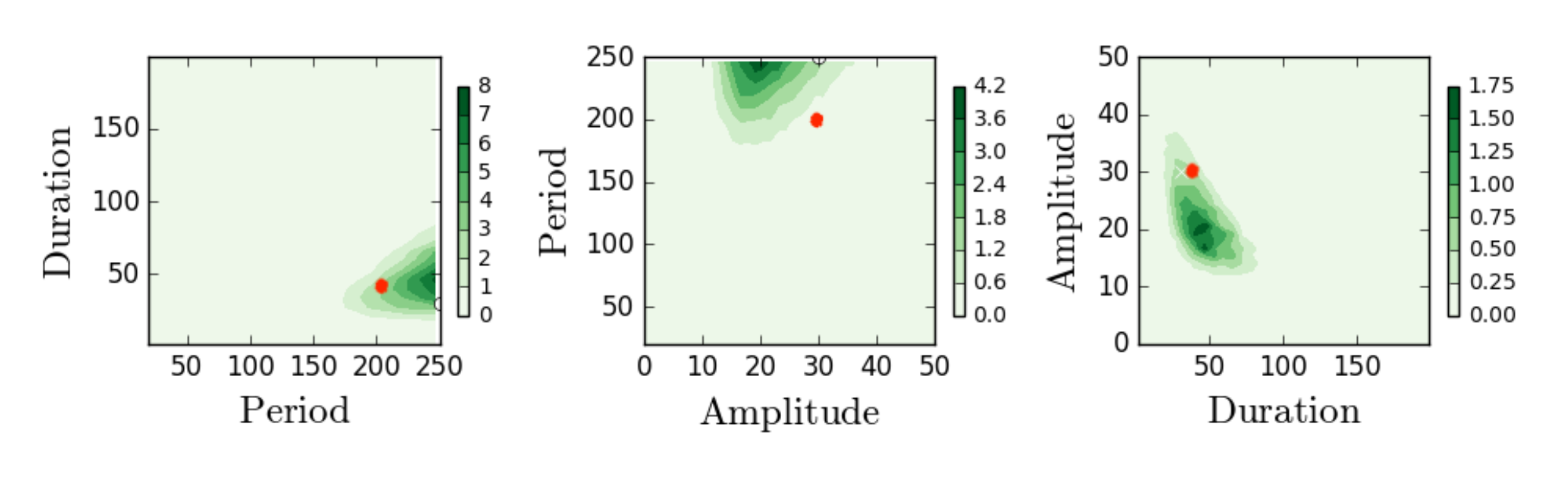} 	% linewidth: shifts the figure to right or left

\includegraphics[width=1 \linewidth]{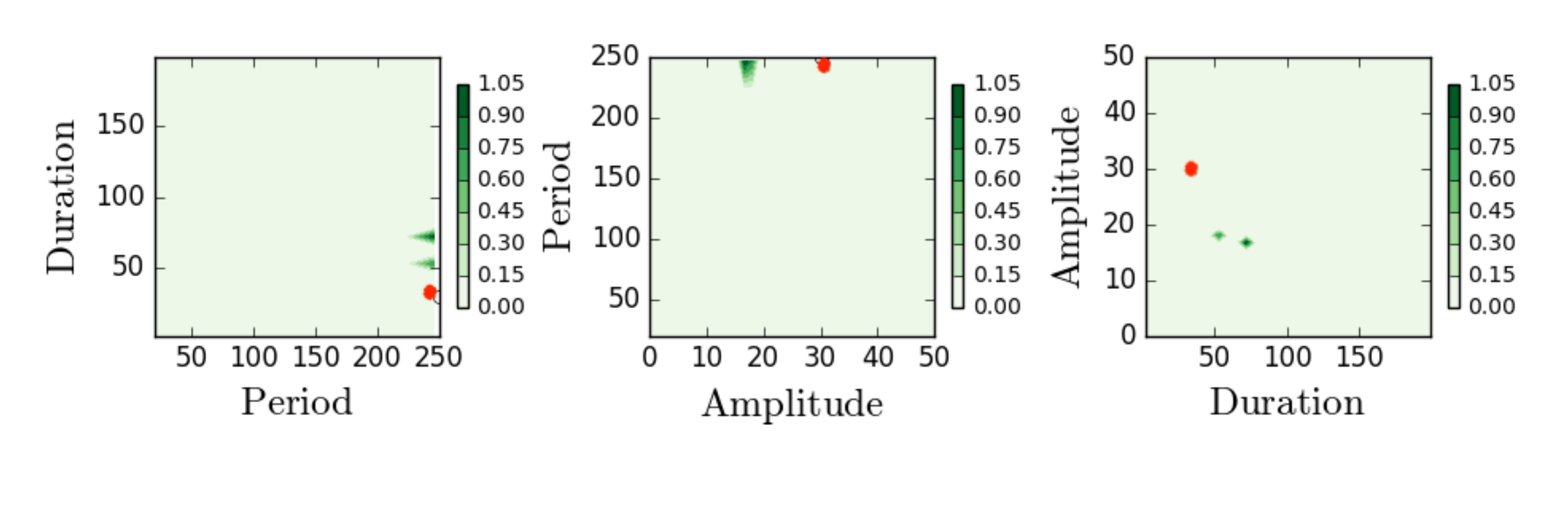}

\caption{2D  contour plots of (D,P), (P,A) and (A, D) from left to right. Parameter space is limited to the W12 range, i.e. $D\leq 200 $, $P\leq 250$, $A<50$ and assuming stellar metallicity of $1 Z_{\odot}$. Top and bottom subplots are contour plots of $M\leq 10^{7}\ M_\odot$ and $10^7<M\leq10^8\  M_{\odot}$ respectively. The maximum likelihood values are in good agreements with the W12 predictions, shown as red circles.}
\label{fig:8}
\end{center}
\end{figure*}

We first test our method by comparing our analysis with that of W12 to see whether we produce the same best-fit values. To do so, we make the same assumptions as W12 (assuming solar metallicity and identical parameter space). Figure \ref{fig:8} shows the  contours of the burst parameters for the two lowest mass ranges in W12, i.e. $M\leq 10^{7}\ M_\odot$ and $10^7<M\leq10^8\ M_{\odot}$.

As can be seen in Figure \ref{fig:8}, the best-fit parameters are in agreement with W12. They report (D, P, A) of  (40, 200, 30) and (30, 250, 30) for $M\leq 10^{7}\ M_\odot$ and $10^7<M\leq10^8\  M_{\odot}$ respectively which is comparable to our results; (46, 250, 20) (Figure \ref{fig:8}, top) and (53, 250, 18) (Figure  \ref{fig:8}, bottom). The slight discrepancy in the duration and amplitude between our method and W12 arises from the different sampling of the parameter space. 
This confirms the credibility of our maximum likelihood approach for determining the best-fit parameters of bursts. The results and the interpretations are discussed in section \ref{sec:review}.

\bibliography{main2} 
\end{document}